\documentclass[pdflatex,sn-mathphys-num]{sn-jnl}

\usepackage{graphicx}%
\usepackage{multirow}%
\usepackage{amsmath,amssymb,amsfonts}%
\usepackage{amsthm}%
\usepackage{mathrsfs}%
\usepackage[title]{appendix}%
\usepackage{xcolor}%
\usepackage{textcomp}%
\usepackage{manyfoot}%
\usepackage{booktabs}%
\usepackage{algorithm}%
\usepackage{algorithmicx}%
\usepackage{algpseudocode}%
\usepackage{listings}%

\usepackage{cleveref}
\usepackage{mathtools}
\usepackage{bm}
\usepackage{bbding}

\newcommand{\R}{\mathbb{R}}
\newcommand{\opdiv}{\operatorname{div}}
\newcommand{\diag}{\operatorname{diag}}
\DeclareMathOperator*{\argmin}{\textrm{argmin}}
\DeclareMathOperator{\rank}{rank}
\DeclareMathOperator{\trace}{trace}
\DeclareMathOperator{\orth}{orth}
\newcommand{\normin}[1]{\lVert #1 \rVert}
\newcommand{\norm}[1]{\left \lVert #1 \right \rVert}
\newcommand{\absin}[1]{\lvert #1 \rvert}
\newcommand{\abs}[1]{\left \lvert #1 \right \rvert}

\theoremstyle{thmstyleone}%
\theoremstyle{thmstyletwo}%

\theoremstyle{thmstylethree}%

\begin{document}

\title[The Filter Echo: 

A General Tool for Filter Visualisation]
{The Filter Echo: 

A General Tool for Filter Visualisation}


\author*[1]{\fnm{Daniel} \sur{Gaa}}\email{gaa@mia.uni-saarland.de}

\author[1]{\fnm{Joachim} \sur{Weickert}}\email{weickert@mia.uni-saarland.de}

\author[1]{\fnm{Iva} \sur{Farag}}\email{farag@mia.uni-saarland.de}
\equalcont{Iva Farag and \"{O}zg\"{u}n \c{C}i\c{c}ek were affiliated with 
the Mathematical Image Analysis Group during the time they worked on their 
student theses.}

\author[1]{\fnm{\"{O}zg\"{u}n} \sur{\c{C}i\c{c}ek}}\email{cicek@mia.uni-saarland.de}
\equalcont{Iva Farag and \"{O}zg\"{u}n \c{C}i\c{c}ek were affiliated with 
the Mathematical Image Analysis Group during the time they worked on their 
student theses.}

\affil[1]{\orgdiv{Mathematical Image Analysis Group, Faculty of 
Mathematics and Computer Science}, 
\orgname{Saarland University},
\orgaddress{\street{Campus E1.7}, \city{66041 Saarbr\"{u}cken}, 
\country{Germany}}}


\abstract{
To select suitable filters for a task or to improve existing filters, a deep 
understanding of their inner workings is vital. Diffusion echoes, which 
are space-adaptive impulse responses, are useful to visualise the effect of 
nonlinear diffusion filters. However, they have received little attention in 
the literature. There may be two reasons for this: Firstly, the concept was 
introduced specifically for diffusion filters, which might appear 
too limited. Secondly, diffusion echoes have large storage requirements, 
which restricts their practicality. This work addresses both problems. 
We introduce the filter echo as a generalisation of the diffusion echo and 
use it for applications beyond adaptive smoothing, such as image inpainting, 
osmosis, and variational optic flow computation. We provide a framework to 
visualise and inspect echoes from various filters with different applications. 
Furthermore, we propose a compression approach for filter echoes, which 
reduces storage requirements by a factor of 20 to 100.
}

\keywords{Diffusion Echo, Impulse Response, Filter Kernel, 
Singular Value Decomposition}



\maketitle

%
%

\section{Introduction}

Even in times where deep neural networks are highly popular, model-based
approaches remain appealing due to their simplicity, transparency, and
mathematical foundation. One of the simplest model-based approaches
are linear shift-invariant (LSI) filters, for which it is well known that
they can be represented by convolutions. Their filter action is not
space-variant and is determined by the so-called \emph{impulse response}.
It describes the filter output of an image with a unit impulse in the 
origin in the discrete case, or a delta peak in the continuous setting.

Since LSI filters may be too limited for various practical applications, 
nonlinear adaptive filters have been introduced. In the case of 
denoising applications, they allow e.g.~to create structure-preserving 
results. Obviously, these highly adaptive filters cannot have a simple 
impulse response that characterises and visualises the filter behaviour
at all locations. By design, their action is space-variant and adapts 
itself to the original image. In some cases, this action is still
transparent: For example, bilateral filtering~\cite{AW95,SB97,TM98} 
and nonlocal (NL) means~\cite{BCM05a} use space-variant adaptive 
filter kernels that are readily available and can thus be visualised 
in a straightforward way.

For filters based on nonlinear partial differential equations
(PDEs) such as nonlinear diffusion filters \cite{PM90,Ha94,We97}, such   
an immediate intuition is not directly available. In 2001, Dam and 
Nielsen introduced the diffusion echo~\cite{DN01} as a means of 
intuitive understanding of diffusion filters. 
The (source) diffusion echo at a given location is obtained by evolving a 
unit impulse according to the given nonlinear diffusion evolution. Thus, 
diffusion echoes can be seen as the space-variant analogue of the impulse 
response~\cite{PM13} of an LSI filter. While the impulse response of 
an LSI filter is space-invariant and does not depend on the input image, 
the diffusion echo differs between locations and changes with the input.

The concept is powerful: Diffusion echoes carry the full information on the 
filtering process. If all echoes are known, they can be used to reconstruct the 
filtered image from the original. More importantly, the echoes are rich in 
information and offer a full understanding of the underlying diffusion process. 
Therefore, they can be used to analyse the filter behaviour in dependence on 
parameters or discretisations, or to investigate changes between different 
scales in the diffusion scale-space~\cite{Ii62}.

Unfortunately, in spite of its obvious merits, the diffusion echo has 
received little attention in the literature so far. This may have two
reasons:

Firstly, the work of Dam and Nielsen does not go beyond diffusion 
filtering. However, other PDE-based filters might also benefit from an 
interpretation in terms of an echo. Furthermore, variational methods 
in image processing and computer vision are naturally connected to
PDEs via their Euler-Lagrange equations or their gradient  
flow~\cite{AK06,GWWB08,HS81,IH81,WTBW05}. This suggests a generalisation 
of the diffusion echo to other filters.

Secondly, an apparent drawback of diffusion echoes are their high storage 
costs. Since the echo may differ from pixel to pixel, an image of size 
$N$ exhibits $N$ echoes of the same size, which means that storage 
requirements grow quadratically in the number of pixels. 
Thus, $4 \cdot 10^{9}$ floating point numbers, which equates to 
16 gigabytes of data, must be stored for an image of size $256 \times 256$.
Therefore, a more compact representation of the diffusion echo is desirable.


\subsection{Our Contribution}

This paper highlights, generalises, and improves the practical applicability 
of the concept of the diffusion echo by addressing both problems. 
Firstly, we propose a general filter echo framework, which subsumes the 
diffusion echo formulation~\cite{DN01}, and show how general smoothing 
filters and additional image processing and computer vision filters can be 
included in the framework. It comprises not only the previously mentioned 
filters, but also inpainting~\cite{Ca88,GWWB08,MHWT+11}, 
osmosis~\cite{WHBV13}, and optic flow~\cite{HS81} models. 
The basic matrix formulation in the framework is of general type and is
commonly used~\cite{Mi13a}.
Secondly, we propose a compression strategy for filter echoes, which mitigates 
their biggest drawback. Our approach is based on probabilistic 
algorithms~{\cite{HMT11,PRT+00,RST10}} for computing a truncated singular value 
decomposition (SVD)~\cite{GL96}. We test this strategy on a number of 
filters and show that we can substantially reduce the storage requirements 
while allowing for a straightforward reconstruction in terms of a single 
matrix-vector multiplication.

\smallskip
The present manuscript builds upon our conference contribution~\cite{GWFC25}, 
in which we review the concept of the diffusion echo~\cite{DN01}, give an 
interpretation in terms of a typical numerical approach, and introduce our 
SVD-based compression approach for the efficient representation of isotropic 
nonlinear diffusion echoes. We extend our conference paper~\cite{GWFC25}  
with the following additional contributions:
\begin{itemize}
    \item[$\bullet$] We introduce a general framework for filter echoes. 
    We show that it subsumes the diffusion echo formulation and cast a 
    number of smoothing filters as well as general filters into the 
    appropriate form.
    \item[$\bullet$] We display the generality and versatility of the filter 
    echo as a tool for visualising filters. We show its usefulness for
    \begin{itemize}
        \item highlighting differences between filters commonly used for the
        same task.
        \item understanding and highlighting which components of filters 
        make them specifically powerful for a given task.
        \item interpreting complex filters in a more intuitive manner.
    \end{itemize}
    \item[$\bullet$] We extend the echo compression experiments 
    from~\cite{GWFC25} to other filters. We show that the case of isotropic 
    nonlinear diffusion with Weickert diffusivity that we considered 
    in~\cite{GWFC25} is of particular difficulty. Our results show that the 
    compression potential of our approach is substantially higher for other 
    diffusion filters. 
\end{itemize}


\subsection{Related Work}

As already mentioned, the idea of the diffusion echo has been introduced
by Dam and Nielsen~\cite{DN01}. 
In the context of inpainting-based image compression, the use of an 
\emph{inpainting echo} has been proposed to optimise the tonal values of the 
stored data~\cite{MHWT+11}. However, it has never been visualised, but has 
been used as a purely algorithmic tool.

The drain diffusion echo~\cite{DN01} can be interpreted as a space-variant 
local filter kernel. Such local kernels have been visualised, for example, for 
the bilateral filter~\cite{TM98}, the nonlocal means filter~\cite{BCM05a} or 
the guided filter~\cite{HST12}. In contrast to diffusion filters, these methods 
explicitly design the shape of the kernels by assigning weights for the local 
averaging. Therefore, the local filter kernels can be retrieved directly.

The diffusion (and filter) echo gives an exact representation of the filtering
process. However, this representation is costly. There exist approaches that
aim to approximate nonlinear diffusion kernels, e.g.~by using adaptively 
shaped Gaussians~\cite{DON03} or by learning space-variant integral 
kernels~\cite{FS97}. However, these approaches do not give quantifiable
approximation results.

Other approaches can be linked to nonlinear diffusion using e.g.~scaling limits
or iterative applications, even though they were not initially designed to 
approximate diffusion processes. Examples are the Nitzberg--Shiota 
filter~\cite{NS92}, the bilateral filter~\cite{AW95,SB97,TM98}, or nonlocal 
linear diffusion scale-spaces~\cite{CWS15}.

To compress filter echoes, we apply a truncated SVD~\cite{GL96} that we compute 
with a common method from probabilistic linear algebra, based on
a randomised singular value decomposition (RSVD)~{\cite{HMT11,PRT+00,RST10}}.
By discarding components from the SVD we can steer the reconstruction error 
in a consistent and quantifiable manner.

Loosely related to our work is the idea of Milanfar~\cite{Mi13}, who uses 
symmetric approximations of filters that he then decomposes with an 
eigendecomposition. The symmetrisation guarantees orthogonality of the 
eigenvectors, which enables the visualisation of the local effect of the 
filter on some exemplary shapes. Another mildly related work deals with
the short time kernel for the Beltrami flow \cite{SKS03}. However,
it uses a kernel that describes the increment in time, while our
filter echo characterises the accumulated action over time.

Previous work by two of the coauthors was done in the form of two student
theses that employ a principal component analysis (PCA)~\cite{Pe01} to 
compress diffusion echoes~\cite{Ba16,Ci14a}. Although this approach is 
related, they work on a subset of the echo data and discard certain echoes, 
adding the additional task of selecting an appropriate subset. We work on the 
full matrix, which frees us from such considerations. This furthermore allows 
us to calculate the RSVD without explicitly computing and storing any echoes 
and without computing explicit matrix-vector multiplications.


\subsection{Paper Organisation}

In \Cref{sec:framework} we present the general filter echo framework, which
allows us to transfer the concepts of the diffusion echo to a broader class of 
widely used image processing and even computer vision filters.
In \Cref{sec:fitting-filters} we show how a large number of different filters
can be formulated in such a way that they fit our framework. This enables us 
to consider meaningful echoes for these filters. In \Cref{sec:compression} we 
introduce our compression framework for the filter echo.
Our experiments are presented in \Cref{sec:experiments}. We first display 
and discuss a number of different filter echoes.
Then we evaluate the compression approach on different diffusion filters, 
showing that we can reduce the storage requirements by a considerable amount, 
making the entire concept more practical. Eventually, we conclude our work in 
\Cref{sec:conclusions}.


\section{The Filter Echo Framework}
\label{sec:framework}

In the following, we introduce our general filter echo framework,
which we define in the discrete setting. Although some of the filters that 
we present are derived in the continuous setting, they need to be discretised 
before being applied to digital images, which makes the discrete setting 
adequate for our considerations.

We define discrete images on a regular pixel grid of size $n_x \times n_y$ 
and stack them into vectors of size $N \coloneqq n_x n_y$. 
In our framework, we consider filters that can be written as the result
of a matrix-vector multiplication between a matrix 
$\bm{S} \in \R^{N \times N}$ and the original image $\bm{f} \in \R^{N}$.
Then the resulting filtered image $\bm{u} \in \R^{N}$ is given by
\begin{equation}
    \label{eq:framework}
    \bm{u} = \bm{S} \, \bm{f}.
\end{equation}
We call $\bm{S}$ the \emph{state transition matrix} of the filter. It is 
important to note that $\bm{S}$ is not restricted to being a fixed 
shift-invariant linear filter, but might depend on the initial image 
$\bm{f}$ in the case of linear space-variant or nonlinear filters. 
$\bm{S}$ is quadratic in the number of pixels and is generally dense. 
It contains the complete information on the filtering process and maps the 
initial image $\bm{f}$ to the filtered solution $\bm{u}$. In the following 
sections, we show that we can express a large number of different filters 
by means of such a matrix. However, storing it or sometimes even computing 
it can be very challenging in practice.

We now introduce the idea of a filter echo in terms of the discrete filter 
\labelcref{eq:framework}. The original idea goes back to Dam and 
Nielsen~\cite{DN01}, who introduced the diffusion echo. However, they restrict
themselves to the continuous setting and do not provide discrete theory or an 
embedding into a general discrete framework. Furthermore, they do not consider
any filters beyond diffusion filters.

Adhering to their definitions~\cite{DN01}, the \emph{source echo} $\bm{s}_i$ at 
location $i$ is the result of filtering a unit impulse $\bm{e}_i$ centred in $i$. 
Using our description of the filter in terms of the state transition matrix 
\labelcref{eq:framework}, we can express the source echo as
\begin{equation}
    \label{eq:source_echo}
    \bm{s}_i = \bm{S} \, \bm{e}_i.
\end{equation}
This shows that \emph{the source echo $\bm{s}_i$ of a filter is given by the 
$i$-th column of its state transition matrix $\bm{S}$}. The source echo 
describes how the grey value data from $f_i$ is distributed by the filtering 
process. It is a space-variant generalisation of the impulse response of an 
LSI filter~\cite{PM13} and can be interpreted as the perspective of the 
``sender''.

Analogously, the \emph{drain echo}~\cite{DN01} shows where the grey values 
$u_j$ originated from. Using the formulation \labelcref{eq:framework}, 
\emph{the drain echo $\bm{d_j}$ is given by the $j$-th row of the state 
transition matrix $\bm{S}$}:
\begin{equation}
    \label{eq:drain_echo}
    \bm{d}_j = \bm{S}^\top \bm{e}_j.
\end{equation}
Therefore, the drain echo corresponds to the local, space-variant filter 
kernel. It can be interpreted as the ``receiver'' perspective.

By definition~\cite{DN01}, the $j$-th component of the source echo in pixel 
$i$ is the same as the $i$-th component of the drain echo in pixel $j$, that 
is $(\bm{s}_i)_j = (\bm{d}_j)_i$. From the matrix-based formulation this is 
straightforward to see.
Since the source and drain echoes correspond to the columns and rows of the 
state transition matrix $\bm{S}$, equality between both echoes at a pixel 
position is only guaranteed for symmetric state transition matrices.

If all echoes are known, they can be used to reconstruct the filtered image 
$\bm{u}$. From \labelcref{eq:framework} it follows that 
\begin{equation}
    \label{eq:source-echo-reco}
    \bm{u} = \bm{S} \, \bm{f} = \sum_{k=1}^{N} f_k \bm{s}_k,
\end{equation}
for source echoes, and that the component-wise reconstruction from drain
echoes is given by
\begin{equation}
    \label{eq:drain-echo-reco}
    u_j = \bm{d}_j^{\top} \bm{f},
\end{equation}
for $j=1, \dots, N$.


\section{Analysing Filters within our Framework}
\label{sec:fitting-filters}

In this section, we first show that our framework subsumes the diffusion echo 
formulation of Dam and Nielsen~\cite{DN01}. Next, we demonstrate that we can 
straightforwardly extend the approach to further smoothing filters. Afterwards, 
we show that other image processing and computer vision algorithms can also be
reformulated such that they exhibit echoes.


\subsection{Echoes for Smoothing Filters}
\label{subsec:smoothing-echo}

We start by considering diffusion processes, which are given by a PDE. We show 
how an exemplary numerical solution strategy can be used to fit the continuous 
process into our discrete framework~\labelcref{eq:framework}. This serves as 
a blueprint for later considerations.

\Cref{eq:framework} is not new in terms of smoothing 
filters~\cite{Mi13a}. Well-established filters, such as, for example, the 
bilateral filter~\cite{AW95,SB97,TM98}, the NL means filter~\cite{BCM05a}
or the guided filter~\cite{HST12} are specifically designed as weighted 
averages. From a filter echo perspective, 
these are straightforward cases, as the echoes are directly obtained from the 
weights. However, note that this does not hold, for example, for the iterated 
bilateral filter~\cite{El02}, which can be expressed in terms of its echoes, 
but whose echoes cannot be retrieved directly. Nevertheless, we also include 
the bilateral filter and NL means into our considerations, since visualisations 
of the echoes are still interesting and instructive.


\subsubsection{The Diffusion Echo}
\label{subsubsec:diff-echo}

We now consider the original idea that is the basis of our work on a
general filter echo: the \emph{diffusion echo}~\cite{DN01}.
Expressing a diffusion process given by a PDE in terms of 
\labelcref{eq:framework} is not straightforward, so we describe the necessary 
steps in the following.

Diffusion filters are typically derived in a continuous setting, on a 
two-dimensional, rectangular domain $\Omega \subset \R^2$, where grey value 
images are defined as mappings from $\Omega$ to $\R$. For the prototype of a 
diffusion evolution~\cite{We97}, we consider the parabolic initial boundary 
value problem
\begin{alignat}{3}
\label{eq:diff-pde} \partial_t u(\bm{x}, t) &= \opdiv(\bm{D} \, \bm{\nabla} 
u(\bm{x}, t))     \quad &\textnormal{for }& 
\bm{x} \in \Omega, \,\,t \in (0, \infty), \\
\label{eq:diff-ic} u(\bm{x}, 0) &= f(\bm{x}) 
    \quad &\textnormal{for }& \bm{x} \in \Omega, \\
\label{eq:diff-bc} \partial_{\bm{n}} u(\bm{x}, t) &= 0 
    \quad &\textnormal{for }& 
        \bm{x} \in \partial \Omega, \,\,t \in (0, \infty).
\end{alignat}
Here, $f \colon \Omega \to \R$ is the initial image at $t=0$, and 
$u \colon \Omega \times [0, \infty) \to \R$ is the evolving image. 
Moreover, $\bm{\nabla} = \left( \partial_x, \partial_y \right)^\top$ is 
the spatial gradient and $\opdiv (\bm{v}) = \partial_x v_1 + \partial_y v_2$ 
is the spatial divergence for $\bm{v} = (v_1, v_2)^{\top}$. 
The matrix $\bm{D}$ is the \emph{diffusion tensor}, a symmetric, positive 
definite matrix of size $2 \times 2$, which steers the smoothing behaviour
of the diffusion process. This diffusion tensor may depend on the evolving
image $u$, in which case it renders the PDE nonlinear. Lastly, $\bm{n}$ is the 
outer normal at the boundary of the image domain $\partial \Omega$.

To solve \labelcref{eq:diff-pde,eq:diff-bc,eq:diff-ic}, we discretise in space 
and time. Space discretisation is performed on a regular pixel grid as 
described in \Cref{sec:framework}.
This yields the semi-discrete (time-continuous and space-discrete) scheme
\begin{align}
    \frac{d \bm{u}(t)}{d t}
    &= \bm{A}(\bm{u}(t)) \bm{u}(t), 
    \label{eq:diff-space-discrete-eq} \\
    \bm{u}(0) &= \bm{f}. \label{eq:diff-space-discrete-bc} 
\end{align}
The matrix $\bm{A}(\bm{u}(t))$ adequately discretises the spatial differential 
operators in \labelcref{eq:diff-pde} and includes the reflecting boundary 
conditions \labelcref{eq:diff-bc}. There are
five conditions (S1)--(S5) that were formulated for such matrices to guarantee
that the process fulfils important theoretical properties, such as uniqueness
of a solution, fulfilment of a maximum-minimum principle or preservation of the 
average grey value. These conditions on $\bm{A}(\bm{u}(t))$ are 
Lipschitz-continuity in $\bm{u}$, symmetry, vanishing row sums, nonnegative 
off-diagonals, and irreducibility~\cite[Chapter 3]{We97}.

Lastly, the time variable $t$ is discretised with time step size $\tau$. 
We use a semi-implicit scheme, where we fix the nonlinearity in each step:
\begin{equation}
    \label{eq:semi-implicit}
    \frac{\bm{u}^{k+1} - \bm{u}^{k}}{\tau} = 
    \bm{A}(\bm{u}^{k}) \, \bm{u}^{k+1}.
\end{equation}
The upper index denotes the current time step $k$. To comply with the 
initial condition \labelcref{eq:diff-space-discrete-bc}, we set 
$\bm{u}^0 = \bm{f}$. In contrast to an explicit scheme, the semi-implicit 
scheme is stable for arbitrary time step sizes $\tau$, allowing us to use 
fewer time steps to compute the solution at a large time $t$. However, it 
requires us to solve a linear system of equations to compute the solution 
$\bm{u}^{k+1}$ from the current solution $\bm{u}^{k}$:
\begin{equation}
    \label{eq:discrete-diff-step-sys}
    \left( \bm{I} - \tau \bm{A}(\bm{u}^{k}) \right) \, \bm{u}^{k+1} = \bm{u}^{k}.
\end{equation}
Writing the solution explicitly yields
\begin{equation}
    \label{eq:discrete-diff-step}
    \bm{u}^{k+1} = 
    \underbrace{\left( \bm{I} - \tau \bm{A}(\bm{u}^{k}) 
    \right)^{-1}}_{\bm{P} (\bm{u}^k, \tau)} \bm{u}^{k}.
\end{equation}
Similarly to the semi-discrete case, there are also conditions (D1)--(D6) 
formulated for the fully discrete scheme. They are continuity, symmetry, 
unit row sums, nonnegativity, irreducibility, and positive diagonal 
entries~\cite[Chapter 4]{We97}. One can show that $\bm{P} (\bm{u}^k, \tau)$
fulfils all of them and that the existence of the inverse in 
\labelcref{eq:discrete-diff-step} is guaranteed for any $\tau>0$, if 
$\bm{A}(\bm{u}^{k})$ satisfies the five semi-discrete conditions 
(S1)--(S5)~\cite[Chapter 4]{We97}. 
Note that although $\left( \bm{I} - \tau \bm{A}(\bm{u}^{k}) \right)$ is usually 
a sparse matrix, its inverse is generally not.

To make the diffusion models comply with our general filter echo framework 
\labelcref{eq:framework}, we write the solution after $n$ steps in terms of 
the original image $\bm{f}$:
\begin{equation}
 \label{eq:discrete-diff-state-transition}
 \bm{u}^{n} = \underbrace{\bm{P} (\bm{u}^{n-1}, \tau) \cdots 
    \bm{P} (\bm{u}^{1}, \tau)\,
    \bm{P}(\bm{f}, \tau)}_{\bm{S}(\bm{f}, \tau, n)} \, \bm{f}.
\end{equation}
The \emph{state transition matrix} $\bm{S}(\bm{f}, \tau, n)$ for a given 
diffusion process then depends on the initial image $\bm{f}$, the time step 
size $\tau$ and the number of steps $n$. These considerations show that the 
iterative solution of the discretised, nonlinear PDE can be expressed by a 
single matrix-vector multiplication with the large and dense state transition 
matrix.

Assuming that the nonnegativity, symmetry, and unit row sum conditions are 
satisfied, the matrices $\bm{P}(\bm{u}^k, \tau)$ are doubly stochastic 
(nonnegative, with unit row and column sums). It is easy to show that the 
product of doubly stochastic matrices and therefore the state transition 
matrix is doubly stochastic as well, which allows for an interpretation of 
drain and source echoes as probability distributions.
This property of the state transition matrix has important consequences: Unit
row sums and nonnegativity imply a discrete maximum-minimum principle, while
unit column sums imply preservation of the average grey 
value~\cite[Chapter~4]{We97}.
Note that symmetry of the state transition matrix is not guaranteed, even
though the individual $\bm{P}(\bm{u}^k, \tau)$ are symmetric. \emph{This 
means that source and drain echoes usually differ.}

For multiplication with the transposed state transition matrix, which is 
needed for drain echo calculation, the symmetric diffusion matrices 
$\bm{P}(\bm{u}^k, \tau)$ are simply applied in reverse order:
\begin{equation}
 \label{eq:discrete-diff-state-transition-transposed}
 \bm{d}_j = \bm{S}(\bm{f}, \tau, n)^\top \, \bm{e}_j 
 = \bm{P}(\bm{f}, \tau) \, \bm{P} (\bm{u}^{1}, \tau) \cdots 
 \bm{P} (\bm{u}^{n-1}, \tau) \, \bm{e}_j.
\end{equation}
In practice, this implies that the results from all steps need to be computed 
and stored, before the drain echoes can be calculated. The source echoes, on 
the other hand, can be computed along with the image evolution.

It is important to note that the considerations about the state transition
matrix are mainly of theoretical nature. In practice, the fully discrete scheme 
is typically solved by solving the linear systems in each step. Although the 
system matrix $\left( \bm{I} - \tau \bm{A}(\bm{u}^{k}) \right)$ is quadratic 
in the number of pixels, it is usually sparse. This can be exploited by using 
iterative solvers, such as the conjugate gradient (CG) method~\cite{Vo03}, 
which only rely on the evaluation of matrix-vector products and do not require 
the explicit formation of the system matrix.
However, this means that it is neither desirable to explicitly compute
$\bm{P} (\bm{u}^k, \tau)$, nor $\bm{S}(\bm{f}, \tau, n)$, and that 
multiplications with them should always be evaluated via the corresponding 
linear systems.

The preceding considerations entail a large number of different diffusion
filters, some of which we will now examine in more detail.


\smallskip
\emph{Homogeneous Linear Diffusion}

Let us first consider the simplest diffusion model, homogeneous linear
diffusion~\cite{Ii62}. It is obtained by setting the diffusion tensor to
the identity, i.e.~$\bm{D} = \bm{I}$, for which the diffusion PDE 
\labelcref{eq:diff-pde} simplifies to 
$\partial_t u = \Delta u = \partial_{xx} u + \partial_{yy} u$. 
It is well known that homogeneous diffusion in an infinite domain is equivalent 
to Gaussian convolution with a Gaussian kernel of standard deviation 
$\sigma = \sqrt{2 t}$~\cite{Co88}. Therefore, the diffusion echoes are 
Gaussian kernels that incorporate the reflecting boundary conditions.


\smallskip
\emph{Isotropic Nonlinear Diffusion}

Space-adaptive, direction-independent (\emph{isotropic}) smoothing behaviour
can be achieved by setting the diffusion tensor to a multiple of the
identity, with the magnitude changing depending on the location. 
A common choice is 
\begin{equation}
    \bm{D}(\bm{\nabla} u) = g \left(\abs{\bm{\nabla} u}^2 \right) \bm{I},
\end{equation}
where $\absin{\cdot}$ denotes the Euclidean norm.
This yields what we call isotropic nonlinear 
diffusion\footnote{Note that the terminology in the literature is not 
consistent. We differentiate between isotropic (direction-independent) and
anisotropic (direction-dependent) diffusion processes.}~\cite{PM90}. The 
scalar function $g$ is known as \emph{diffusivity}~\cite{PM90}. 
It is a positive, decreasing function of the gradient magnitude and locally 
steers the strength of the diffusion activity using a nonlinear feedback
mechanism depending on the evolving image $u$. 
It is common to use a Gaussian-smoothed version of the gradient in the 
diffusivity~\cite{CLMC92}, i.e.~$\bm{\nabla} u_\sigma$, with $\sigma$ being 
the standard deviation of the Gaussian.

Popular diffusivity functions include the Charbonnier 
diffusivity~\cite{CBAB97}:
\begin{equation}
   \label{eq:ch-diff}
   g_{ch}(s^2) 
   = \frac{1}{\sqrt{1 + s^2 / \lambda^2}},
\end{equation}
the rational Perona--Malik diffusivity~\cite{PM90}:
\begin{equation}
   \label{eq:pm-diff}
   g_{pm}(s^2) 
   = \frac{1}{1 + s^2 / \lambda^2},
\end{equation}
or the Weickert diffusivity~\cite{We97}:
\begin{equation}
   \label{eq:weickert-diff}
   g_{we}(s^2) 
   = 
   \begin{cases}
        1.0, & \text{if} \; s^2 = 0, \\
        1.0 - \exp\left({\frac{-3.3148}{s^8/\lambda^8}}\right), & \text{else},
   \end{cases}
\end{equation}
with a positive contrast parameter $\lambda$.


\smallskip
\emph{Edge-Enhancing Anisotropic Nonlinear Diffusion}

While the previous filter offers a space-variant smoothing behaviour,
for certain applications, however, an \emph{anisotropic} smoothing~\cite{We94e} 
may be preferable. 
A popular representative of anisotropic diffusion filters is the so-called
edge-enhancing diffusion (EED)~\cite{We94e}. EED reduces the diffusion
activity across image edges while still allowing full diffusive smoothing 
along them. To this end, one selects the diffusion tensor as
\begin{equation}
    \bm{D}(\bm{\nabla} u_{\sigma}) = g \left(\vert \bm{\nabla} u_\sigma 
    \vert^2\right)
    \frac{\bm{\nabla} u_\sigma}{\lvert \bm{\nabla} u_\sigma \rvert} 
    \left(\frac{\bm{\nabla} u_\sigma}{\lvert \bm{\nabla} u_\sigma \rvert}
    \right)^{\top} + 1 \, \frac{\bm{\nabla} u_\sigma^{\perp}}
    {\lvert \bm{\nabla} u_\sigma^{\perp} \rvert}
    \left(\frac{\bm{\nabla} u_\sigma^{\perp}}
    {\lvert \bm{\nabla} u_\sigma^{\perp} \rvert} \right)^{\top}.
\end{equation}
This formulation implies that $\bm{D}$ has an eigenvector 
$\bm{v}_1 = \frac{\bm{\nabla} u_\sigma}{\lvert \bm{\nabla} u_\sigma \rvert}$
with corresponding eigenvalue 
$\lambda_1 = g \left(\vert \bm{\nabla} u_\sigma \vert^2\right)$, which inhibits
smoothing across strong image edges. The second eigenvector is orthogonal
to the first with eigenvalue $\lambda_2 = 1$, which leads to full smoothing
along the image edge.
It should be noted that while discretising the continuous PDE works similarly 
to isotropic nolinear diffusion, anisotropic PDEs require additional, careful 
considerations, since a standard discretisation in space may violate the 
nonnegativity condition~\cite{We97}. As a remedy, $L^2$-stable 
discretisations have been proposed~\cite{WWW13}, which bound the occurring 
over- and undershoots.


\subsubsection{The Bilateral Filter Echo}
\label{subsubsec:bilateral-echo}

Bilateral filtering averages pixels using weights that depend on 
\emph{closeness} (distance in the domain) and \emph{similarity} 
(distance in the co-domain)~\cite{AW95,SB97,TM98}.
In the setting with a regular pixel grid and a grey value image, which we 
consider, the closeness can be measured by the Euclidean distance and 
the similarity by the absolute difference of the grey values.

The discrete bilateral filter is then expressed for all $i=1, \dots, N$ by 
the weighted average
\begin{equation}
    \label{eq:bilateral}
    u_i = \frac{\sum_{j=1}^{N} g \left( \abs{f_{i} - f_{j}} \right) 
    w \left( \abs{\bm{x}_i - \bm{x}_j} \right) f_j}
    {\sum_{j=1}^{N} g \left( \abs{f_{i} - f_{j}} \right) 
    w \left( \abs{\bm{x}_i - \bm{x}_j} \right)} = \sum_{j=1}^{N} p_{i, j} f_j,
\end{equation}
with
\begin{equation}
    p_{i, j} = \frac{g \left( \abs{f_{i} - f_{j}} \right) 
    w \left( \abs{\bm{x}_i - \bm{x}_j} \right)}
    {\sum_{\ell=1}^{N} g \left( \abs{f_{i} - f_{\ell}} \right) 
    w \left( \abs{\bm{x}_i - \bm{x}_\ell} \right)}.
\end{equation}
For the tonal and spatial weighting functions $g$ and $w$, we consider
Gaussians with standard deviations $\sigma_t$ and $\sigma_s$.
The final weight $p_{i,j}$ describes the influence of pixel $j$ of
the input image $\bm{f}$ on pixel $i$ of the filter output ${\bm u}$.

With this notation, we can rewrite \labelcref{eq:bilateral} as
\begin{equation}
    \bm{u} = \bm{P} (\bm{f}) \bm{f},
\end{equation}
with a state transition matrix $\bm{P}(\bm{f}) = \left( p_{i,j} \right)$,
which for fixed weighting functions depends only on the initial image $\bm{f}$. 
We note that $u_i$ is given as a convex combination of all $f_j$, with 
$\sum_j p_{i,j} = 1$ for all $i$ and $p_{i,j} \geq 0$ for all $i,j$. 
Therefore, $\bm{P}$ has unit row sums, and all its entries are non-negative, 
which implies that the bilateral filter fulfils a discrete maximum-minimum 
principle~\cite[Chapter~4]{We97}. Such matrices are also called 
\emph{row stochastic}. However, due to the normalisation, $\bm{P}$ is 
nonsymmetric and may not have unit column sums, such that the average grey 
value may not be preserved~\cite[Chapter~4]{We97}.
Since $\bm{P}$ is only row stochastic, we can interpret only its drain
echoes as probability distributions.


\subsubsection{The Nonlocal Means Echo}

Nonlocal (NL) means~\cite{BCM05a} is related to bilateral filtering,
in the sense that it is also given as a weighted average. However, its
weights are based on the similarity between two patches $\mathcal{N}_i$ 
and $\mathcal{N}_j$, which can be defined as neighbourhoods (e.g.~disk-shaped) 
around the pixels $i$ and $j$.
The similarity between two patches with $\left\vert \mathcal{N} \right\vert$ 
pixels is given by the Euclidean distance between the corresponding grey value 
vectors $\bm{f}(\mathcal{N}_i) \in \R^{\vert \mathcal{N} \vert}$ and 
$\bm{f}(\mathcal{N}_j) \in \R^{\vert \mathcal{N} \vert}$. As for bilateral
filtering, a weighting function $g$ is applied. We consider a Gaussian 
with standard deviation $\sigma$. This yields the filtered result
\begin{equation}
    \label{eq:nlmeans}
    u_i = \frac{\sum_{j=1}^{N} g \left(\abs{\bm{f}(\mathcal{N}_i) 
    - \bm{f}(\mathcal{N}_j)} \right) f_j}
    {\sum_{j=1}^{N} g \left(\abs{\bm{f}(\mathcal{N}_i) 
    - \bm{f}(\mathcal{N}_j)} \right) } = \sum_{j=1}^{N} p_{i, j} f_j.
\end{equation} 
with
\begin{equation}
    p_{i, j}= \frac{g \left(\abs{\bm{f}(\mathcal{N}_i) 
    - \bm{f}(\mathcal{N}_j)} \right))}
    {\sum_{\ell=1}^{N} g \left(\abs{\bm{f}(\mathcal{N}_i) 
    - \bm{f}(\mathcal{N}_\ell)} \right)}.
\end{equation}

The matrix $\bm{P}(\bm{f}) = (p_{i, j})$ is the state transition matrix of 
the process. Its properties are the same as for bilateral filtering, 
i.e.~it depends on $\bm{f}$, is row stochastic and nonsymmetric. Satisfaction 
of a maximum-minimum principle follows, while the preservation of the average 
grey value is not guaranteed.


\subsection{The Inpainting Echo}
\label{subsec:inpainting-echo}

We now diverge from smoothing filters and extend our filter echo framework to 
(sparse) PDE-based inpainting. The \emph{inpainting echo} has originally been 
introduced as a tool to efficiently solve the tonal optimisation problem in 
image compression~\cite{MHWT+11}, and we will first review the basic theory 
behind it. 

In the inpainting setting, we assume that image data is known only on a 
subset $K$ of the image domain $\Omega$. This subset is called \emph{mask}. 
We aim at reconstructing the image in the unknown areas 
$\Omega \setminus K$ by solving a PDE using an inpainting operator $L$, which 
may or may not depend on the evolving image $u$. 

We consider the elliptic inpainting formulation, which is instructive in terms 
of inpainting echoes, since some of their properties can be directly derived 
from the form of the state transition matrix.
The full elliptic boundary value problem is given by
\begin{alignat}{3}
\label{eq:inp-pde-ell} L u(\bm{x}) &= 0 
    \quad &\textnormal{for }& \bm{x} \in \Omega \setminus K, \\
\label{eq:inp-init-ell}  u(\bm{x}) &= f(\bm{x}) 
    \quad &\textnormal{for }& \bm{x} \in K, \\
\label{eq:inp-bound-ell} \partial_{\bm{n}} u(\bm{x}) &= 0 
    \quad &\textnormal{for }& \bm{x} \in \partial \Omega.
\end{alignat}
Examples of suitable inpainting operators include homogeneous diffusion
inpainting ($L u = \Delta u$)~\cite{Ca88}, nonlinear diffusion inpainting
($L u = \opdiv (g \, \bm{\nabla} u)$) \cite{SPMEWB14}, or EED inpainting 
($L u = \opdiv (\bm{D} \, \bm{\nabla} u)$)~\cite{GWWB08,WW06}.

By introducing a mask function $c \colon \Omega \to \{0, 1\}$ that takes 
the value $1$ at mask locations and $0$ elsewhere, we can merge 
\labelcref{eq:inp-pde-ell,eq:inp-init-ell} into a single equation:
\begin{equation}
\label{eq:inp-joint}
    c(\bm{x}) (u(\bm{x}) - f(\bm{x})) - (1 - c(\bm{x})) L u(\bm{x}) = 0.
\end{equation}

Discretisation yields
\begin{equation}
    \bm{C} (\bm{u} - \bm{f}) - (\bm{I} - \bm{C}) \bm{L}(\bm{u})\bm{u} = \bm{0},
\end{equation}
where $\bm{I}$ is the identity matrix, $\bm{C} = \diag(\bm{c})$ is a diagonal 
matrix with $\bm{c}$ on the diagonal, and $\bm{L}$ is the discrete analogue of 
the differential operator with reflecting boundary conditions.
Reordering the terms yields a system of equations, which, depending on the
differential operator, is linear or nonlinear:
\begin{equation}
    \label{eq:inp-system}
     \left( \bm{C} - (\bm{I} - \bm{C}) \bm{L}(\bm{u}) \right) \, \bm{u} 
     = \bm{C} \bm{f}.
\end{equation}

For a linear differential operator, we have $\bm{L}(\bm{u}) = \bm{L}$, and 
a linear system is solved to obtain the inpainted solution $\bm{u}$:
\begin{equation}
\label{eq:inp-sol}
    \bm{u} = \underbrace{\left(\bm{C} - (\bm{I} - \bm{C}) \bm{L} \right)^{-1} 
    \bm{C}}_{\bm{S}(\bm{c})} \, \bm{f}.
\end{equation}
For homogeneous diffusion inpainting, the existence of the inverse is 
guaranteed as long as we have at least one mask pixel~\cite{MBWF11}. For a 
fixed linear differential operator, the nonsymmetric state transition matrix 
is uniquely determined by the mask configuration. 

In the nonlinear case, we can solve \labelcref{eq:inp-system} using the 
Ka\v{c}anov method~\cite{Ka59}. This means that we fix the nonlinearity in 
each step according to the current solution, leading to a number of linearised 
problems of similar form as \labelcref{eq:inp-system}:
\begin{equation}
    \left(\bm{C} - (\bm{I} - \bm{C}) \bm{L}
    (\bm{u}^k) \right) \bm{u}^{k+1} = \bm{C} \, \bm{f},
\end{equation}
where the upper index denotes the iteration step and where the initialisation 
$\bm{u}^0$ satisfies $\bm{C} \bm{u}^0 = \bm{C}\bm{f}$.
Note that the final solution $\bm{u}^n$ after $n$ steps depends only on 
$\bm{L}(\bm{u}^{n-1})$, which means that we do not have to store all 
intermediate solutions to compute the echoes.
Assuming an invertible system matrix, we obtain  
\begin{equation}
    \label{eq:kacanov-sol}
    \bm{u}^{n} = \underbrace{\left(\bm{C} - (\bm{I} - \bm{C}) \bm{L} 
    (\bm{u}^{n-1}) \right)^{-1} \bm{C}}_{\bm{S}(\bm{c}, \bm{f}\vert_{\bm{c}})} 
    \, \bm{f}.
\end{equation}
In addition to the mask, the state transition matrix $\bm{S}$ now also depends 
on the grey values at the mask locations $\bm{f}\vert_{\bm{c}}$.

\Cref{eq:kacanov-sol} is very insightful w.r.t.\ the echo configuration. 
Due to the right multiplication with $\bm{C}$ in the state transition matrix, 
the source echoes in nonmask pixels vanish entirely, which reflects that only 
the pixel values at mask locations have an influence on the final inpainting 
result. Accordingly, the drain echoes only take on nonzero values at mask 
locations. It was shown that for homogeneous diffusion inpainting, $\bm{S}$ has 
nonnegative entries and unit row sums, so the inpainting process satisfies a 
maximum-minimum principle~\cite{MBWF11}. 
Numerically, we compute the elliptic inpainting solution for nonlinear 
operators as the steady state of a parabolic evolution with automatic time
step size adaptation.

\begin{figure}[!bt]
\centering
\tabcolsep3pt
\begin{tabular}{ccc}
\includegraphics[width=0.3\linewidth]
{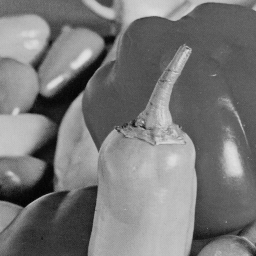} & 
\includegraphics[width=0.3\linewidth]
{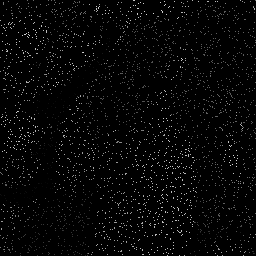} & 
\includegraphics[width=0.3\linewidth]
{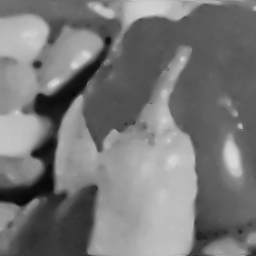} \\
\small{(a) original image} & \small{(b) stored data} &
\small{(c) EED reconstruction}
\end{tabular}
\caption{\label{fig:eed-inp} Inpainting example using the test image
\emph{peppers}. $5\%$ randomly selected pixels are stored and the rest is
discarded. Then the image is reconstructed using EED inpainting with the
Charbonnier diffusivity ($\lambda=0.8$, $\sigma=1.0$).}
\end{figure}

\medskip
\Cref{fig:eed-inp} shows an example of sparse inpainting. Only $5\%$ of the
pixels in \emph{peppers} are stored and the rest is discarded. Then, EED 
inpainting with the parabolic formulation is used to restore the missing 
information. Even though the stored pixels were chosen randomly, the inpainting 
reconstructs the image with an adequate quality and even restores the edges. 
Note that quality can be drastically improved, if the stored data is 
optimised~\cite{SPMEWB14}.


\subsection{The Osmosis Echo}
\label{subsec:osmosis-echo}

In the next step, we consider linear osmosis modelled by a nonsymmetric 
drift-diffusion process~\cite{WHBV13}, which can create details and leads to
nonconstant steady states. Applications range from shadow removal or 
image cloning \cite{WHBV13} to image stitching \cite{BPW23}.

Unlike in the diffusion case, we assume that we are given a \emph{positive}
initial image $f \colon \Omega \to \R^N_+$ and a \emph{drift vector field} 
$\bm{d} \colon \Omega \to \R^2$ that steers the osmosis process.
A family of filtered images $\{u(\bm{x}, t) \mid t \geq 0\}$ is obtained 
by solving
\begin{alignat}{3}
\label{eq:os-pde} \partial_t u(\bm{x}, t) 
    &= \opdiv \left(\bm{\nabla} u(\bm{x}, t) 
    - \bm{d}(\bm{x}) u(\bm{x}, t) \right) 
    \quad &\textnormal{for }& \bm{x} \in \Omega, \,\,t \in (0, \infty) \\
\label{eq:os-ic} u(\bm{x}, 0) &= f(\bm{x}) 
    \quad &\textnormal{for }& \bm{x} \in \Omega, \\
\label{eq:os-bc} 0 &= \bm{n}^{\top} (\bm{\nabla} u(\bm{x}, t) 
- \bm{d}(\bm{x}) u(\bm{x}, t)) 
    \quad &\textnormal{for }& 
        \bm{x} \in \partial \Omega, \,\,t \in (0, \infty).
\end{alignat}
Typically one is only interested in the steady state ($t \to \infty$) of this 
evolution.

Like in the diffusion case, we discretise our image on a regular pixel 
grid. Again, an adequate discretisation of the differential operators leads 
to a semi-discrete problem of a form similar to 
\labelcref{eq:diff-space-discrete-eq}. However, the drift term renders the 
matrix $\bm{A}$ non-symmetric. The given process is linear, so $\bm{A}$ does 
not depend on $\bm{u}$. As in the diffusion case, there are conditions 
(SLO1)--(SLO3) on $\bm{A}$ in the semi-discrete case. These are zero column 
sums, nonnegativity of the off-diagonal entries, and irreducibility. More 
details can be found in~\cite{VHWS13}. 

For the time discretisation, we use an implicit scheme, which satisfies 
the relevant conditions (DLO1)--(DLO4) (unit column sums, nonnegativity, 
irreducibility, positive diagonal entries)~\cite{VHWS13} on the nonsymmetric 
matrix $\bm{P} \in \R^{N \times N}$ of the fully discrete scheme
\begin{align}
    \bm{u}^0 &= \bm{f}, \\
    \bm{u}^{k+1} &= \underbrace{\left( \bm{I} 
    - \tau \bm{A}\right)^{-1}}_{\bm{P} (\tau)} \bm{u}^{k},
\end{align}
for all $\tau > 0$. We solve the linear systems with BiCGSTAB~\cite{Vo03}. 
We can write the result after $n$ steps of size $\tau$ as
\begin{equation}
    \bm{u}^{n} = \bm{P}^n(\tau) \, \bm{f} = \bm{S}(\tau, n) \, \bm{f}.
\end{equation}

Since the iteration matrix $\bm{P}$ has unit column sums~\cite{VHWS13}, 
the same holds for $\bm{S}$. Unit row sums are not guaranteed. 
Osmosis therefore preserves the average grey value, but does not
fulfil a maximum-minimum principle.

In this work, we only consider \emph{the compatible case}: 
Setting the drift vector field to $\bm{d} = \bm{\nabla} (\ln {v})$, with some 
guidance image $v$, linear osmosis converges to 
$u = \frac{\mu_{v}}{\mu_{f}} v$, with $\mu_{v}$ and $\mu_{f}$ being the average 
grey values of $v$ and $f$, respectively~\cite{WHBV13}. However, note that 
the preceding analysis about the state transition matrix holds irrespective 
of the specific drift vector field.

\begin{figure}[!tb]
\centering
\tabcolsep3pt
\begin{tabular}{cccc}
\small{(a) initial image} & \small{(b) guidance image} & & \\[0.5mm]
\includegraphics[width=0.23\linewidth]
{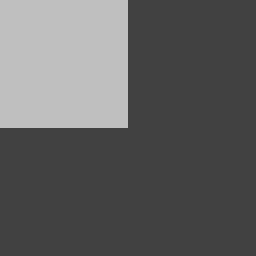} & 
\includegraphics[width=0.23\linewidth]
{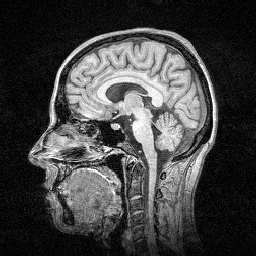} & 
& \\[3pt]
\includegraphics[width=0.23\linewidth]
{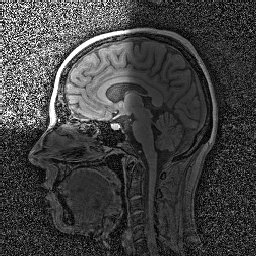} & 
\includegraphics[width=0.23\linewidth]
{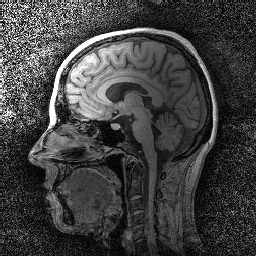} & 
\includegraphics[width=0.23\linewidth]
{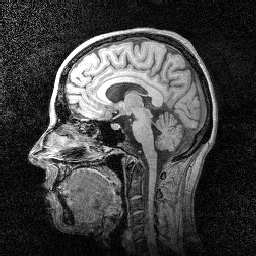} & 
\includegraphics[width=0.23\linewidth]
{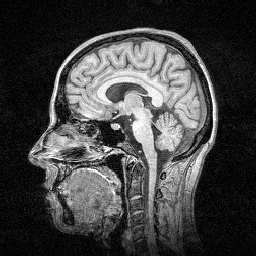} \\
\small{(c) $T=25$} & \small{(d) $T=250$} & 
\small{(e) $T=2500$} & \small{(f) $T=500000$}
\end{tabular}
\caption{\label{fig:osm-evol} Osmosis evolution in the compatible case with 
semi-implicit scheme visualised at different stopping times $T$. 
Test image \emph{square} is the initial image, and \emph{head} the guidance 
image. The initial image is rescaled, such that its average grey value matches
that of the guidance image. For $t \to \infty$ the guidance image is recovered.}
\end{figure}

In \Cref{fig:osm-evol} we show the results of an osmosis process with the 
\emph{head} test image as a guidance image, from which we derive the drift
vector field, and a simple initial image. We rescale the initial image
such that its average grey value matches that of the guidance image, so
the evolving image converges exactly to the guidance image.


\medskip
Although osmosis is a powerful process, its steady-state echoes are 
surprisingly simple. To show that, we make use of results of Proposition 1 
in~\cite{VHWS13}, which are based on the eigendecomposition 
$\bm{P} = \bm{Q} \bm{\Lambda} \bm{Q}^{-1}$, with 
$\bm{Q} = \left[\bm{q}_1 \cdots \bm{q}_N\right]$.
The results, which mostly follow from the Perron--Frobenius theory for 
nonnegative matrices (see, e.g.~\cite{HJ90}), state that $\bm{P}$ has a 
simple eigenvalue $\lambda_1=1$ and that all other eigenvalues are strictly 
smaller in absolute value. Furthermore, iterative application of $\bm{P}$ 
yields the steady state $\bm{v} \in \R^{N}_{+}$, which is the eigenvector 
$\bm{q}_1$ to the eigenvalue $\lambda_1=1$~\cite{VHWS13}.
Lastly, due to the column sum property, the eigenvector of $\bm{P}^{\top}$ 
that corresponds to $\lambda_1=1$ is a constant vector~\cite{VHWS13}.

Repeated application of $\bm{P}$ yields the state transition matrix $\bm{S}$ 
of the steady state. Since its eigenvalues are given by powers of the 
eigenvalues of $\bm{P}$, we can conclude that it has a single nonzero 
eigenvalue $\lambda_1=1$. Thus, $\bm{S}$ is a rank-$1$ matrix, given by the 
outer product of $\bm{q}_1$ and the corresponding row of $\bm{Q}^{-1}$.
Since $\bm{P}^{\top} = \left( \bm{Q} \bm{\Lambda} \bm{Q}^{-1} \right)^{\top}
= (\bm{Q}^{-\top} \bm{\Lambda} \bm{Q}^{T})$, the column of $\bm{Q}^{-\top}$ 
corresponding to $\lambda_1$ (which is the corresponding eigenvector of 
$\bm{P}^{\top}$) is constant, so the corresponding row of $\bm{Q}^{-1}$ is 
also constant.
It follows that all columns of the steady-state state transition matrix 
are the same. They are given by the eigenvector $\bm{q}_1$ of 
$\bm{P}$, which is the steady-state solution of the process.

Therefore, the steady-state drain echoes are constant, with intensity 
corresponding to the intensity of the respective pixel of the steady-state 
image divided by the sum of the entries of the initial image. Furthermore, 
all source echoes are given by the same version of the steady-state solution,
i.e.~the rescaled guidance image, again divided by the sum of the entries
of the initial image to satisfy the unit column sum condition.


\subsection{The Optic Flow Echo}
\label{subsec:optic-flow-echo}

Lastly, we take our ideas a step further by considering the echoes of 
a computer vision model: variational optic flow. It is widely acknowledged 
that variational optic flow models gain their power through the filling-in 
effect of the regulariser~\cite{BWS05}, which transports information to the 
areas where the flow is difficult to determine. In fact, Demetz et 
al.~\cite{DWBZ12} show that certain linear variational optic flow models can 
be interpreted as Whittaker--Tikhonov regularisation of the normal flow,
where the Euclidean norm in the regularisation term is replaced by some
norm that is specific to the optic flow constraint.
As we shall see, we can use a similar formulation to produce reasonable optic 
flow echoes with a slightly restructured formulation of 
\labelcref{eq:framework}.

Optic flow models~\cite{HS81} aim to estimate the motion between two 
subsequent frames of an image sequence at times $t$ and $t+1$. They produce 
a flow field $\bm{w} = \left(u, v\right)^\top$, where $u$ and $v$ describe 
the motion along the $x$- and $y$-directions, respectively. 

A basic assumption is that the grey values of a point in the first frame and 
of the corresponding point in the second frame are the same~\cite{HS81}, 
which can be expressed as $f(x, y, t) = f(x+u, y+v, t+1)$.
Linearisation via a first-order Taylor expansion yields the constraint
\begin{equation}
    \label{eq:of-constraint}
    f_x u + f_y v + f_t = \bm{\nabla} f^{\top} \bm{w} + f_t = 0,
\end{equation}
where $f_x \coloneqq \partial_x f$ is the partial derivative of $f$ w.r.t.\ 
$x$, and $\bm{\nabla} = \left(\partial_x,  \partial_y \right)^\top$ is the
spatial gradient.

We see that arbitrary components orthogonal to $\bm{\nabla} f$ can be added
to the flow field $\bm{w}$ without violating the optic flow constraint
\labelcref{eq:of-constraint}. This is called \emph{aperture problem}.
It follows that \labelcref{eq:of-constraint} can only determine the component 
parallel to $\bm{\nabla} f$, which is called the \emph{normal flow}:
\begin{equation}
\begin{split}
    \label{eq:normal-flow-1}
    \begin{pmatrix}
        u_n \\
        v_n
    \end{pmatrix} 
    &= \bm{w}^{\top} \frac{\bm{\nabla} f}{\abs{\bm{\nabla} f}}
    \frac{\bm{\nabla} f}{\abs{\bm{\nabla} f}} \\
    &= \left( f_x u + f_y v \right)\frac{\bm{\nabla} f}
    {\abs{\bm{\nabla} f}^2}.
\end{split}
\end{equation}
From \labelcref{eq:of-constraint} we have $f_x u + f_y v = - f_t$, so
the normal flow is given by
\begin{equation}
    \label{eq:normal-flow-2}
    \bm{w}_n =
    \begin{pmatrix}
        u_n \\
        v_n
    \end{pmatrix} 
    = - f_t \frac{\bm{\nabla} f}{\abs{\bm{\nabla} f}^2}
    = - \frac{1}{\abs{\bm{\nabla} f}^2}
    \begin{pmatrix}
        f_x f_t \\
        f_y f_t
    \end{pmatrix},
\end{equation}
and its regularised version~\cite{DWBZ12}, which avoids singularities for
vanishing $\bm{\nabla} f$, by
\begin{equation}
    \bm{w}_n = \frac{-f_t}{\abs{\bm{\nabla} f}^2 + \epsilon^2} \bm{\nabla} f,
\end{equation}
where $\epsilon > 0$ is a small constant.
Since the optic flow constraint is insufficient to compute a dense flow field,
variational optic flow models include additional smoothness constraints in 
terms of a regularisation term~\cite{WS00b}. 

We consider two linear models as examples: 
The Horn--Schunck model~\cite{HS81} and the anisotropic, image-driven  
Nagel--Enkelmann model~\cite{NE86}. For both of them, the minimisation process 
leads to linear PDEs.

\smallskip
A suitable energy functional is given by
\begin{equation}
    E(u, v) = \int_\Omega \left(f_x u + f_y v + f_t\right)^2 
    + \alpha V(\bm{\nabla} f, \bm{\nabla} u, \bm{\nabla} v) \, dx \, dy,
\end{equation}
with some regularisation term $V$ and a regularisation parameter $\alpha>0$, 
and the minimising flow field is found in terms of its the Euler--Lagrange 
equations, which are given by diffusion-like PDEs of the following form:
\begin{align}
    0 &= f_x^2 u + f_x f_y v + f_x f_t - \alpha \opdiv (\bm{D}(\bm{\nabla}f) \, u), 
    \label{eq:el-of-u}\\
    0 &= f_x f_y u + f_y^2 v + f_y f_t - \alpha \opdiv (\bm{D}(\bm{\nabla}f) \, v).
    \label{eq:el-of-v}
\end{align}

The Horn--Schunck model~\cite{HS81} uses $\bm{D}(\bm{\nabla}f) = \bm{I}$,
while the Nagel--Enkelmann model~\cite{NE86} employs 
$\bm{D}(\bm{\nabla}f) = \left( \bm{\nabla} f^{\perp} 
(\bm{\nabla} f^{\perp})^{\top} + \lambda^2 \bm{I}\right)/(f_x^2 + f_y^2 + 2 \lambda^2)$. 
It only allows smoothing of the flow field in the direction orthogonal to the local 
image gradient and avoids smoothing across image discontinuities.

Rewriting \labelcref{eq:el-of-u,eq:el-of-v} as a vector-valued equation yields
\begin{equation}
    \begin{pmatrix}
        f_x^2 u + f_x f_y v - \alpha \, \opdiv (\bm{D}(\bm{\nabla}f) \, u) \\
        f_x f_y u + f_y^2 v - \alpha \, \opdiv (\bm{D}(\bm{\nabla}f) \, v)
    \end{pmatrix}
    = - \begin{pmatrix}
        f_x f_t \\
        f_y f_t
    \end{pmatrix}.
\end{equation}

For the discretisation, we assume that the partial derivatives of $f$ are
computed with suitable finite difference approximations, and the results are
stacked into vectors $\bm{f}_x$, $\bm{f}_y$, and $\bm{f}_t$. Furthermore,
multiplications between these vectors are to be understood in a component-wise 
manner. Assuming an appropriate discretisation of the differential operators
that we denote by $\bm{L}(\bm{f})$, we get the following linear system of 
equations:
\begin{equation}
    {\underbrace{
    \begin{pmatrix}
        \diag(\bm{f}_x^2) - \alpha \bm{L}(\bm{f}) & \diag(\bm{f}_x \bm{f}_y) \\
        \diag(\bm{f}_x \bm{f}_y) & \diag(\bm{f}_y^2) - \alpha \bm{L}(\bm{f})
    \end{pmatrix}}_{\bm{B}}}
    \begin{pmatrix}
        \bm{u} \\
        \bm{v}
    \end{pmatrix} =
    \begin{pmatrix}
        -\bm{f}_x \bm{f}_t \\
        -\bm{f}_y \bm{f}_t
    \end{pmatrix}.
\end{equation}
For an invertible matrix $\bm{B}$, which exists in all nontrivial scenarios,
we can write the explicit solution formally as 
\begin{equation}
    \begin{pmatrix}
        \bm{u} \\
        \bm{v}
    \end{pmatrix} =
    \bm{B}^{-1}
    \begin{pmatrix}
        -\bm{f}_x \bm{f}_t \\
        -\bm{f}_y \bm{f}_t
    \end{pmatrix}.
\end{equation}

\begin{figure}[!tb]
\centering
\tabcolsep3pt
\begin{tabular}{cccc}
\includegraphics[width=0.25\linewidth]
{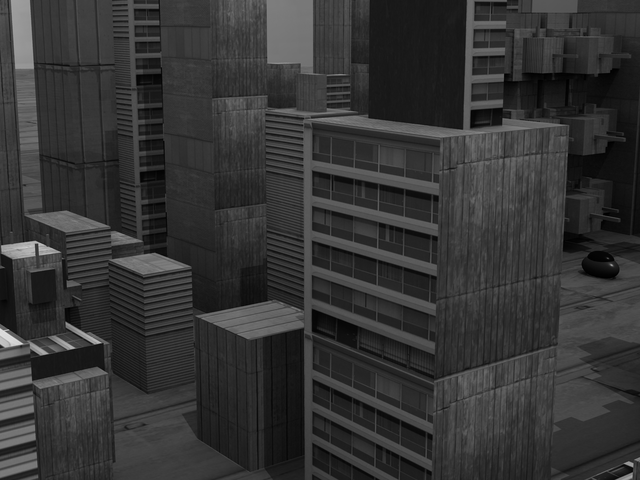} & 
\includegraphics[width=0.25\linewidth]
{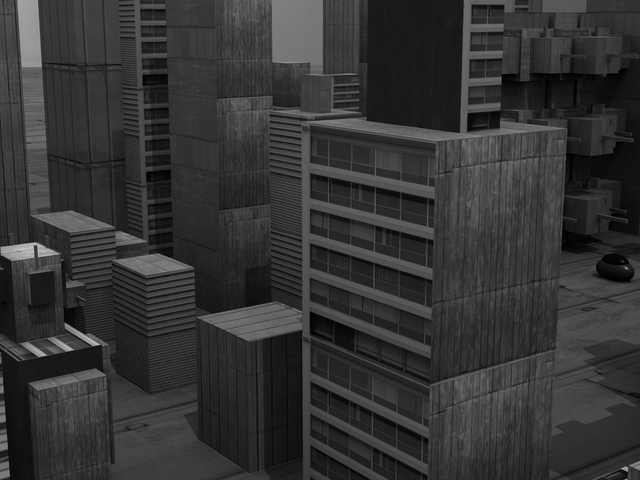} & 
\includegraphics[width=0.25\linewidth]
{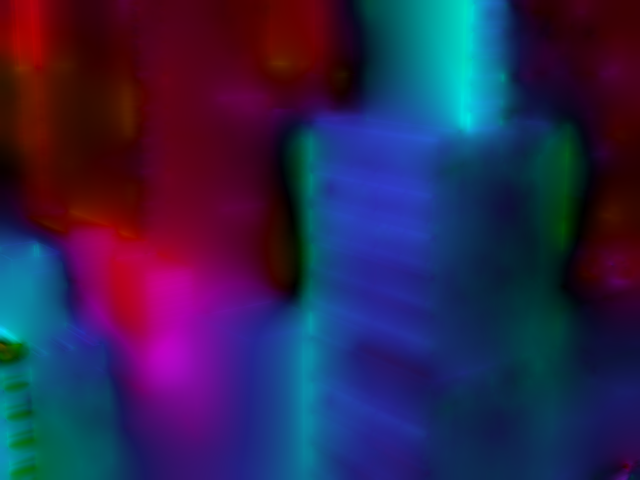} & 
\includegraphics[width=0.1875\linewidth]
{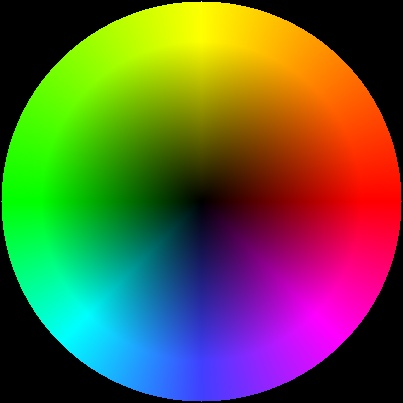} \\
\small{(a) frame 12} & \small{(b) frame 13} &
\small{(c) computed flow} & \small{(d) colour code}
\end{tabular}
\caption{\label{fig:hs-of} Estimation of the optic flow field (colour-coded)
using the Horn--Schunck method ($\alpha=10000$) on frame 12 and frame 13 of 
the \emph{Urban} test sequence from the Middlebury flow data set~\cite{BSLR+11} 
and the used colour code.}
\end{figure}

\Cref{fig:hs-of} shows an example of a flow field computed with the 
Horn--Schunck model~\cite{HS81}. The flow field displays the estimated motion 
between two frames of a test sequence from the Middlebury 
dataset~\cite{BSLR+11}. 


Following the ideas of Demetz et al.~\cite{DWBZ12}, we write the optic flow 
computation in terms of the regularised normal flow, which is directly 
computable from the given frames:
\begin{equation}
\begin{split}
    \begin{pmatrix}
        \bm{u} \\
        \bm{v}
    \end{pmatrix} 
    &=  
    \bm{B}^{-1} \,
    \diag \left(\bm{f}_x^2 + \bm{f}_y^2 + \epsilon^2 \right)
    \diag \left(\bm{f}_x^2 + \bm{f}_y^2 + \epsilon^2 \right)^{-1}
    \begin{pmatrix}
        -\bm{f}_x \bm{f}_t \\
        -\bm{f}_y \bm{f}_t
    \end{pmatrix} \\
    &=  
    \underbrace{
    \bm{B}^{-1} \,
    \diag \left(\bm{f}_x^2 + \bm{f}_y^2 + \epsilon^2 \right)}_{
    \bm{S}(\bm{f}_x, \bm{f}_y, \alpha, \epsilon)}    
    \begin{pmatrix}
        \bm{u}_n \\
        \bm{v}_n
    \end{pmatrix}.
\end{split}
\end{equation}

The state transition matrix $\bm{S}(\bm{f}_x, \bm{f}_y, \alpha, \epsilon)$ 
maps the normal flow to the resulting flow field and depends on the image 
derivatives, as well as on the regularisation parameters $\alpha$ and 
$\epsilon$. While $\bm{B}^{-1}$ is symmetric, multiplication with the diagonal 
matrix renders the state transition matrix $\bm{S}$ nonsymmetric. 
Note that $\bm{S} \in \R^{2N \times 2N}$, so we get $2 N$ echoes of size $2 N$.
In each position, there is an echo for the two components of the optic flow
vector.

This formulation generalises our filter echo framework, since we are 
not filtering the original images, but rather the (regularised) normal flow. 
The matrix $\bm{B}$ contains terms that result from discretisations of
diffusion operators, suggesting that the sparse normal flow experiences some 
sort of sophisticated smoothing. This highlights the relation of variational 
optic flow models to PDE-based smoothing and inpainting methods.
In \Cref{sec:experiments}, we confirm that interpretation by visualising 
source echoes, which provide the information flow of the normal flow data.


\section{Our Echo Compression Framework}
\label{sec:compression}

We have seen that the filter echoes correspond to the columns and rows of
the state transition matrix. An image with $N$ pixels therefore has $N$ source 
echoes and $N$ drain echoes of size $N$ each ($2N$ for optic flow), which 
constitutes a huge amount of data and makes storage costly. 
Furthermore, computing echoes at the time they are needed might not be feasible.
Consequently, an effective compression approach for echoes is desirable. 
The goal is to come up with an alternative, more efficient representation, 
which further allows one to reconstruct the echoes in a short time.

As the source and drain echoes of a filter constitute its state transition 
matrix, it is natural to consider matrix approximation approaches for the 
echo compression task. 

First attempts at such a compression have been made in some unpublished student
theses by coauthors of this work~\cite{Ba16,Ci14a}. The basic idea of these
approaches is to select a subset of echoes and perform a principal component 
analysis (PCA)~\cite{Pe01}.

Our goal is to work with the full, original state transition matrices,
which frees us of the task to (empirically) decide on the most important
echoes and has certain algorithmic advantages. To this end, we use the 
truncated singular value decomposition~\cite{GL96} in conjunction with a 
randomised linear algebra approach to compute it, which does not require the 
explicit formation of the state transition matrix~\cite{HMT11,PRT+00,RST10}. 
We will provide details in the following subsections.


\subsection{The Singular Value Decomposition}
\label{subsec:svd}

The singular value decomposition (SVD)~\cite{GL96} of a matrix 
$\bm{A} \in \R^{m \times n}$ is given by 
\begin{equation}
    \bm{A} = \bm{U} \bm{\Sigma} \bm{V}^{\top}.
\end{equation}
The nonnegative singular values 
$\sigma_1 \geq \sigma_2 \geq \dots \geq \sigma_p \geq 0$, 
with $p=\min(m, n)$, are the diagonal entries of the diagonal matrix 
$\bm{\Sigma} \in \R^{m \times n}$. 
$\bm{U} \in \R^{m \times m}$ and $\bm{V} \in \R^{n \times n}$ are orthogonal 
matrices, containing the left and right singular vectors as columns.

If the rank $r$ of the given matrix $\bm{A}$ is smaller than $p$, $\bm{A}$
has only $r$ nonzero singular values, and we can write
\begin{equation}
    \bm{A} = \bm{U}_r \bm{\Sigma}_r \bm{V}_r^{\top} 
    = \sum_{i=1}^{r} \sigma_i u_i v_i^{\top}.
\end{equation}
Here, the matrix $\bm{\Sigma}_r \in \R^{r \times r}$ only contains the first
$r$ singular values, and the matrices $\bm{U}_r \in \R^{m \times r}$ and 
$\bm{V}_r \in \R^{n \times r}$ are obtained by discarding the last
$m-r$ and $n-r$ columns of $\bm{U}$ and $\bm{V}$. 
This representation is often called \emph{compact SVD} in contrast to 
the \emph{full SVD} introduced above.


\subsubsection{Truncated Singular Value Decomposition}

The fundamental Eckart-Young theorem~\cite{EY36} states that the matrix 
$\bm{B}^{\ast}_k$ of rank $k<r$, which minimises the approximation error to a 
given matrix $\bm{A}$ in the Frobenius norm, is given by the rank-$k$ 
truncated SVD of $\bm{A}$:
\begin{equation}
	\argmin_{\bm{B} \in \R^{m \times n}, \text{ rank}(\bm{B})=k} 
	\norm{\bm{A} - \bm{B}}_\text{F}
	= \sum_{i=1}^{k} \sigma_i \bm{u}_i \bm{v}_i^\top 
    = \bm{U}_k \bm{\Sigma}_k \bm{V}_k^{\top} \eqqcolon \bm{B_k^\ast},
\end{equation}
It furthermore quantifies the approximation error via the discarded singular
values:
\begin{equation}
    \label{eq:trunc-svd-error}
	\norm{\bm{A} - \bm{B_k^\ast}}_\text{F} 
	= \norm{\sum_{i=k+1}^{r} \sigma_i \bm{u}_i \bm{v}_i^\top}_\text{F}
	= \sqrt{\sum_{i=k+1}^{r} \sigma_i^2}.
\end{equation}


\subsubsection{Randomised Singular Value Decomposition}

To efficiently compute the truncated SVD, we use a probabilistic 
approximation, known as \emph{randomised singular value decomposition 
(RSVD)}~\cite{HMT11,PRT+00} or 
\emph{randomised subspace iteration (RSI)}~\cite{HMT11,RST10}.
These methods compute the truncated SVD of $\bm{A} \in \R^{m \times n}$ with 
$m \geq n$ in two steps. First, they project $\bm{A}$ onto the column space 
of a matrix $\bm{Q} \in \R^{m \times k}$, $k < r = \rank(\bm{A})$, with 
orthogonal columns: 
\begin{equation}
\label{eq:projection}
    \widehat{\bm{A}} = \bm{Q} \bm{Q}^{\top} \bm{A}.
\end{equation}
Then they compute the compact SVD of $\widehat{\bm{A}}$, which only has 
$k$ nonzero singular values. This is done efficiently by first computing the 
SVD $\bm{Q}^{\top} \bm{A} = \overline{\bm{U}}_k \bm{\Sigma}_k \bm{V}_k$.
The desired SVD $\widehat{\bm{A}} = \bm{U}_k \bm{\Sigma}_k \bm{V}_k$
with $\bm{U}_k \in \R^{m \times k}$, $\bm{\Sigma}_k \in \R^{k \times k}$ 
and $\bm{V}_k \in \R^{n \times k}$ can then be retrieved via
$\bm{U}_k = \bm{Q} \overline{\bm{U}}_k$.

The error of this approximation depends on the error 
$\normin{\widehat{\bm{A}} - \bm{A}}$, so the main task is to appropriately 
determine $\bm{Q}$. This part of the algorithms is also known as the 
\emph{randomised rangefinder}~\cite{HMT11}, as the span of the columns
of $\bm{Q}$ should cover most of the range of $\bm{A}$.

To compute $\bm{Q}$, the given matrix is applied (once or multiple times) to 
a standard Gaussian random matrix $\bm{G} \in \R^{n \times k}$, with columns 
$\bm{g}_k \sim \mathcal{N}(\bm{0}, \bm{I})$. The result is eventually
orthogonalised via a QR decomposition~\cite{GL96}:
\begin{equation}
\label{eq:rangefinder}
    \bm{Q} = \orth\left((\bm{A} \bm{A}^{\top})^{q-1} \bm{A} \bm{G}\right).
\end{equation}
Choosing a larger parameter $q > 0$ for the exponent can improve the 
approximation quality for matrices with a slowly decaying singular value 
spectrum~\cite{HMT11}. Furthermore, to increase numerical stability, it is 
advisable to perform an orthogonalisation $\orth(\cdot)$ after each application 
of the matrix. The standard \emph{randomised singular value decomposition 
(RSVD)}~\cite{HMT11,PRT+00} applies the rangefinder with parameter $q=1$, 
while for $q>1$ the method is known as \emph{randomised subspace iteration 
(RSI)}~\cite{HMT11,RST10}.

Recent advances, known as \emph{randomised block Krylov methods}~\cite{TW23},
use the entire Krylov space generated through repeated matrix multiplications
to construct $\bm{Q}$. While this enhances the quality of the approximation, 
it also increases the memory requirements.

Applying these methods as presented here requires us to specify the target rank 
$k$. In practice, one typically introduces an oversampling parameter $\ell$, 
creates a matrix $\bm{Q} \in \R^{m \times (k+\ell)}$, and eventually discards 
the last $\ell$ singular values and vectors. This further increases the 
approximation quality, especially for slowly decaying spectra. In practice, 
it may suffice to choose the parameter as small as $\ell=10$~\cite{HMT11}.
The excellent work of Halko et al.~\cite{HMT11} provides theoretical bounds
on the approximation quality in dependence of the model parameters (i.e.~$k$,
$q$ and $\ell$). While we assume that the rank $k$ is fixed, it is worth
noting that one can also iteratively increase the size of the matrix $\bm{Q}$ 
and thus the rank of the final SVD, using a pre-selected target threshold for
$\normin{\widehat{\bm{A}} - \bm{A}}$~\cite{HMT11}.


\subsection{The Echo Compression Approach}
\label{subsec:compression-approach}

Our objective is to find an efficient representation $\widehat{\bm{S}}$ of the 
state transition matrix $\bm{S}$ of the considered filtering process. This 
representation  should minimise the Frobenius norm 
$\normin{\bm{S} - \widehat{\bm{S}}}_\text{F}$. This corresponds to a 
minimisation of the mean squared error (MSE) between the compressed and 
original versions of the source and drain echoes, which is a natural error 
measure if one considers each echo as an individual image.

To this end, we compute a truncated SVD of the state transition matrix 
$\bm{S} \in \R^{N \times N}$ by means of the randomised subspace iteration. 
Within the algorithm, matrix-matrix multiplications with the state transition 
matrix $\bm{S}$ need to be computed. Recall that $\bm{S}$ is dense and 
quadratic in the number of pixels. Fortunately, we can avoid constructing it 
explicitly by evaluating the matrix-matrix multiplications by applying the 
respective filter to each of the columns. For example, for diffusion filters 
this corresponds to computing a diffusion evolution. This approach allows us 
to compress filter echoes without ever needing to explicitly compute even a 
single echo, let alone the full state transition matrix.

Instead of $N^2$ floats for the full state transition matrix, the truncated SVD 
only requires $2Nk$ floats to store the matrices $\bm{U}_k \in \R^{N \times k}$ 
and $\bm{V}_k \bm{\Sigma}_k \in \R^{N \times k}$. This means that the 
representation is more efficient if $k<N/2$. We show in our evaluation in
\Cref{sec:compression-results} that we can typically select
a very small $k$ without deteriorating the quality of the echoes, so the 
compression gain can be substantial.

Let us briefly discuss the computational burden of our compression approach.
The advantage of randomised methods for SVD computation is that the matrix 
decompositions are calculated on small matrices, and can be implemented using 
dedicated linear algebra packages. The bottleneck of the algorithm are the 
numerous matrix-vector multiplications with $\bm{S}$ and 
$\bm{S}^{\top}$~\cite{MT20}.
The algorithm presented in \Cref{subsec:svd} requires $2q$ multiplications
with $\bm{S}$ or $\bm{S}^{\top}$ (one for the projection and $2q-1$ in the 
rangefinder). The matrices are of size $N \times (k+\ell)$. This results in
$2q(k+\ell)$ diffusion evolutions. A na\"{i}ve computation of all 
echoes would require $N$ of those. Therefore, although not an initial 
objective of our work, the approach may even decrease the computational burden. 
In \Cref{sec:compression-results}, we see that, generally, $2q(k+\ell) < N$ 
in our experiments.

Reconstruction of the echoes from the SVD representation is done by a 
single matrix-vector multiplication:
\begin{equation}
    \bm{s}_i^n 
    = \bm{U}_k \bm{\Sigma}_k \bm{V}_k^{\top} \bm{e_i} 
    = \bm{U}_k  \left[ (\bm{V}_k \bm{\Sigma}_k)^{\top} \right]_i,
\end{equation}
and
\begin{equation}
    \bm{d}_i^n 
    = \left(\bm{U}_k \bm{\Sigma}_k \bm{V}_k^{\top}\right)^{\top} \bm{e_i}  
    = \bm{V}_k \bm{\Sigma}_k \bm{U}_k^{\top} \bm{e_i}
    = \bm{V}_k \bm{\Sigma}_k \left[\bm{U}_k^{\top}\right]_i,
\end{equation}
where $\left[ \bm{A} \right]_i$ denotes the $i$-th column of the matrix 
$\bm{A}$. This implies that the smaller the rank $k$ 
(i.e.~the more compressed the data), the faster the reconstruction.

We test our approach on a number of different diffusion filters and provide 
results and discussions in \Cref{sec:compression-results}.


\section{Experiments}
\label{sec:experiments}


\subsection{Echo Visualisation}
\label{subsec:echo-visualisation}

In the first part of the experiments, we demonstrate the visualisation
qualities of the filter echo. 
We show that it can be beneficial for comparing similar filters and that it 
can be used to display the strengths or to understand the subtle details of 
complex filters. Furthermore, we show that if an adequate model is selected, 
it can also be used for tasks that may not directly come to mind.


\subsubsection{Nonlinear Diffusion, Bilateral Filtering and NL Means Echoes}

Isotropic nonlinear diffusion~\cite{PM90}, bilateral 
filtering~\cite{AW95,SB97,TM98} and nonlocal means~\cite{BCM05a} are 
prevalent classical smoothing filters that can be used for image denoising.
Relations between nonlinear diffusion and bilateral filtering have been 
established in different works (see e.g.~\cite{BCM06}, \cite{DW06}, \cite{SKB01}, 
or the survey paper~\cite{PKTD09} and the references therein) and bilateral 
filtering and NL means are similar in spirit, as they both explicitly model
a weighted averaging of the image (see \Cref{subsec:smoothing-echo}). 
Therefore, we use them as examples to show how the filter echo can be used to 
visualise and emphasise the differences between similar filters.

Tomasi and Manduchi~\cite{TM98} visualise the local kernel (drain echo) for
bilateral filtering at a single exemplary artificial step edge to better 
understand the weight computation of the filter. Buades et al.~\cite{BCM05a} 
display NL means kernels for specific image features and diffusion echoes 
have been displayed by Dam and Nielsen~\cite{DN01}.

\begin{figure}[!tb]
\centering
\tabcolsep3pt
\fboxsep0pt
\begin{tabular}{cccc}
\fbox{\includegraphics[width=0.23\linewidth]
{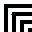}} & 
\includegraphics[width=0.23\linewidth]
{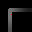} & 
\includegraphics[width=0.23\linewidth]
{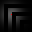} &
\includegraphics[width=0.23\linewidth]
{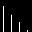} \\
\small{(a) original} &
\small{(b) NLD echo} &
\small{(c) BIL echo} & 
\small{(d) NLM echo} 
\end{tabular}
\caption{\label{fig:nld-bil-nlm} 
Drain echo comparison for different smoothing filters.
(a) original image.
(b) nonlinear diffusion with Weickert diffusivity 
(NLD, $t=150$, $\lambda=0.3$, $\sigma=0.0$) echo. 
(c) bilateral filtering (BIL, $\sigma_t = 30$, $\sigma_s = 10$) echo.
(d) NL means (NLM, patch radius $3$, $\sigma=10$) echo.
The echo location is marked by the red dot.
The three echoes are rescaled jointly, i.e.~the largest echo value 
among all three echoes is mapped to 255.
The nonlinear diffusion echo uses only data from the same segment. Bilateral 
filtering also includes data from tonally similar, unconnected segments, but 
reduces weights for distant pixels. Nonlocal means uses information from 
the entire image, if the local neighbourhood is similar.}
\end{figure}

We use our filter echo framework to visualise and compare local filter 
kernels or drain echoes for all of them on an instructive test case in 
\Cref{fig:nld-bil-nlm}, which sheds light on the different philosophies 
behind the approaches. The parameters are selected such that all edges are
preserved and the filtered images are identical to the original.

A discrete solution to nonlinear diffusion iteratively applies small, local
filters, which are reduced at edges in order to avoid blurring. We see in 
the example that although the resulting kernel can become large 
in its extent, it stops at image edges, as long as their contrast is large 
enough. Therefore, the drain echo is given by a connected segment.
Bilateral filters are capable of ``jumping'' across discontinuities. If a 
segment of similar brightness is spatially separated, it can still have
nozero weights. However, the spatial weighting function decreases for distant 
pixels. 
For NL means, such a spatial weight does not exist. If the search window is
not restricted, it finds similar patches within the entire image, and 
therefore it acts truly global. The patch similarity, which is computed to 
determine the weights, ensures that only pixels with similar local structures 
are considered. For the drain echo in the test image \emph{corners} in 
\Cref{fig:nld-bil-nlm}, this means that only pixels at the right border of 
vertical white stripes are assigned a weight that is notably larger than zero. 


\subsubsection{Nonlinear Diffusion Echoes}
\label{subsubsec:exp-segmentation}

It is known that for appropriate diffusivities and parameters, nonlinear
diffusion can create segmentation-like results~\cite{We97}. The Weickert 
diffusivity with a long stopping time and an appropriate contrast parameter 
is the correct choice for the task~\cite{We97}. However, individual 
segments still have to be extracted from the filtered image. At this point, 
diffusion echoes are an option. They can be used to retrieve and 
identify individual segments~\cite{Ci14a,Je14}. To this end, one simply has 
to extract a source echo, which is located within the segment of interest.

\begin{figure}[!tb]
\centering
\tabcolsep3pt
\begin{tabular}{cccc}
\includegraphics[width=0.23\linewidth]
{images/head.png} & 
\includegraphics[width=0.23\linewidth]
{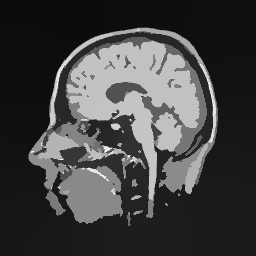} & 
\includegraphics[width=0.23\linewidth]
{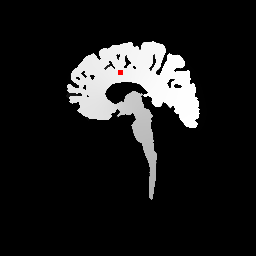} & 
\includegraphics[width=0.23\linewidth]
{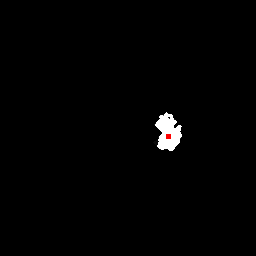} \\
\small{(a) original} &
\small{(b) filtered} &
\small{(c) echo in $(120, 72)$} &
\small{(d) echo in $(168, 136)$}
\end{tabular}
\caption{\label{fig:nld-segmentation} 
Nonlinear diffusion echoes for image segmentation. (a) original image. 
(b) filtered by nonlinear diffusion with the Weickert diffusivity
($t=15000$, $\lambda=5.0$, $\sigma=0.5$).
(c) source echo in $(120, 72)$.
(d) source echo in $(168, 136)$.
The echo locations are marked in red.
The diffusivity creates a segmentation-like result. By computing the source
echo of a pixel in a segment we extract the segment from the filtered result.}
\end{figure}

In \Cref{fig:nld-segmentation} we show an example of such a segmentation.
The \emph{head} test image is smoothed with isotropic nonlinear diffusion with 
parameters selected such that only high-contrast edges are preserved.
The segments of the cartoon-like result are then extracted using source
echoes. We display two source echoes that correspond to distinct segments.
Note how the segments match the structures in the filtered image and are 
adequate representations of semantically relevant structures in the original 
image.


\subsubsection{Sparse Anisotropic Diffusion Inpainting Echoes}

We now visualise some source echoes that help us understand why anisotropic 
diffusion performs so well for sparse inpainting~\cite{GWWB08,SPMEWB14}. 
The source inpainting echoes describe how 
the known data from the mask pixels is distributed to the unknown areas to 
fill in the missing areas. They directly display the information flow
and are the suitable echo choice for inpainting processes.

The goal of our experiments is to show the power of anisotropic EED inpainting
given only sparse data points. We use the Charbonnier diffusivity~\cite{CBAB97}, 
which is commonly used in diffusion-based inpainting~\cite{GWWB08,SPMEWB14},
and employ the parabolic inpainting scheme with a semi-implicit discretisation. 
The linear systems are solved with the conjugate gradient method.
We perform shape completion experiments, inspired by~\cite{SPMEWB14}, which
show how EED inpainting is able to propagate edge information, thanks to
its anisotropy and the Gaussian pre-smoothing in the structure tensor 
computation. By displaying the corresponding source echoes, we get a 
deeper understanding of this information propagation.

\begin{figure}[!tb]
\centering
\tabcolsep3pt
\fboxsep0pt
\begin{tabular}{cccc}
\includegraphics[width=0.23\linewidth]
{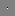} & 
\fbox{\includegraphics[width=0.23\linewidth]
{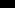}} & 
\fbox{\includegraphics[width=0.23\linewidth]
{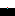}} & 
\fbox{\includegraphics[width=0.23\linewidth]
{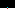}} \\
\small{(a) masked input} &
\small{(b) inpainted} &
\small{(c) source echo} & 
\small{(d) source echo}
\end{tabular}
\caption{\label{fig:dipole-inp} 
Inpainting of the \emph{dipole} test image. (a) original image, the grey 
areas are unknown. (b) inpainted result using EED inpainting ($\lambda=0.01$,
$\sigma=0.1$). (c), (d) source echoes of the two mask pixels. 
The red dot marks the echo location, the cyan dots mark the other mask pixels.
EED spreads the information from the mask pixel to the entire half-plane, creating 
a sharp edge and filling in the image domain.}
\end{figure}

Firstly, we consider the \emph{dipole} experiment~\cite{SPMEWB14} in
\Cref{fig:dipole-inp}. We see that EED is able to create a sharp edge, which 
separates the image into the two desired half-planes. The echoes
of the two mask pixels show that the data is propagated only inside the
corresponding half-plane and that there is no data flow across the boundary.


\begin{figure}[!tb]
\centering
\tabcolsep3pt
\fboxsep0pt
\begin{tabular}{cccc}
\includegraphics[width=0.23\linewidth]
{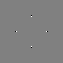} & 
\includegraphics[width=0.23\linewidth]
{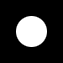} & 
\includegraphics[width=0.23\linewidth]
{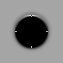} & 
\includegraphics[width=0.23\linewidth]
{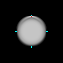} \\
\small{(a) masked input} &
\small{(b) inpainted} &
\small{(c) source echo} & 
\small{(d) source echo}
\end{tabular}
\caption{\label{fig:quadpole-inp} 
Inpainting of the \emph{four dipoles} test image. (a) original image, the grey 
areas are unknown. (b) inpainted result using EED inpainting ($\lambda=0.01$,
$\sigma=1.5$).
(c), (d) source echoes of two mask pixels. 
The red dot marks the echo location, the cyan dots mark the other mask pixels.
EED creates a sharp disk.}
\end{figure}

The \emph{four dipoles} experiment~\cite{SPMEWB14} in \Cref{fig:quadpole-inp} 
builds upon these results. EED is able to reconstruct a white disk with sharp 
boundaries from only eight mask points. The echoes show that each of the 
mask pixels has a truly global influence on the result. The discontinuity 
between the foreground and the background is perfectly respected by the 
information flow. 


\begin{figure}[!tb]
\centering
\tabcolsep3pt
\begin{tabular}{cccc}
\small{(a) original} &
\small{(b) HD} & 
\small{(c) NLD} & 
\small{(d) EED} \\ [0.5mm]
\includegraphics[width=0.23\linewidth]
{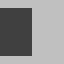} & 
\includegraphics[width=0.23\linewidth]
{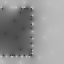} & 
\includegraphics[width=0.23\linewidth]
{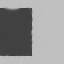} & 
\includegraphics[width=0.23\linewidth]
{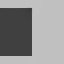} \\[3pt]
\includegraphics[width=0.23\linewidth]
{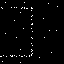} & 
\includegraphics[width=0.23\linewidth]
{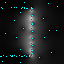} & 
\includegraphics[width=0.23\linewidth]
{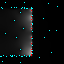} & 
\includegraphics[width=0.23\linewidth]
{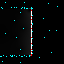} \\
\small{(e) mask} &
\small{(f) cumul. echo} &
\small{(g) cumul. echo} &
\small{(h) cumul. echo}
\end{tabular}
\caption{\label{fig:inp-edge-sq}
Inpainting results of the test image \emph{rectangle} using 
(b) homogeneous diffusion (HD) inpainting, 
(c) isotropic nonlinear diffusion (NLD) inpainting ($\lambda=0.1$) with 
Charbonnier diffusivity, 
(d) EED inpainting ($\lambda=0.1$, $\sigma=0.5$), and corresponding cumulative 
echoes along a prominent image edge. The mask pixels corresponding to 
echoes are marked in red, and the other mask pixels in cyan.
The three cumulative echoes are rescaled jointly.
The echoes show how EED is able to connect edges even if the mask density
is low and the mask locations are suboptimal.}
\end{figure}

In \Cref{fig:inp-edge-sq} we compare EED to other diffusion-based inpainting 
strategies. The \emph{rectangle} experiment shows how the strengths of EED 
allow it to accurately reconstruct edges from limited data, which does not 
have to be as perfectly aligned as in the \emph{dipole} and \emph{four dipoles} 
experiments. We place mask pixels at staggered locations on both sides of the 
edges (see \Cref{fig:inp-edge-sq} e). We then consider all echoes that are 
located left (i.e.~in the dark segment) of the vertical edge and display them 
in a single image. We call this a \emph{cumulative echo}. It shows how the 
filters behave along an edge of the image. We see that homogeneous diffusion 
inpainting spreads the data equally. However, in contrast to homogeneous 
diffusion smoothing (\Cref{subsubsec:diff-echo}), its echo is space-variant: 
It is affected by other mask pixels. 
This allows this simple inpainting method to perform well if the mask locations 
are optimised properly~\cite{MHWT+11}. However, in this case, the imperfect 
mask placement and low density lead to poor edge reconstruction. The cumulative 
echo highlights the undesired data propagation across the edge. 
Isotropic nonlinear diffusion, on the other hand, reduces the smoothing near 
edges. Thus, the grey values do not bleed as much across the edge. However, it 
is still unable to properly connect the edge, as the smoothing is reduced in 
an isotropic way, meaning that also the smoothing along the edge is reduced. 
This is reflected by the cumulative echo. EED mitigates this. The cumulative 
echo shows how the data from the mask pixels is propagated along the edge, 
leading to a sharp reconstruction. Note that there is also some data 
propagation inside the rectangle.


\begin{figure}[!tb]
\centering
\tabcolsep3pt
\begin{tabular}{cccc}
\small{(a) original} &
\small{(b) HD} & 
\small{(c) NLD} & 
\small{(d) EED} \\[0.5mm]
\includegraphics[width=0.23\linewidth]
{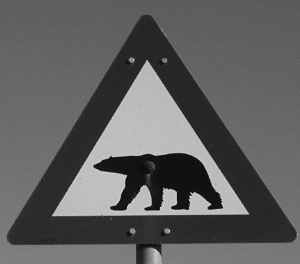} & 
\includegraphics[width=0.23\linewidth]
{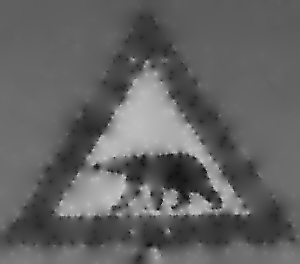} & 
\includegraphics[width=0.23\linewidth]
{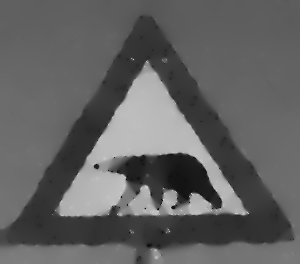} & 
\includegraphics[width=0.23\linewidth]
{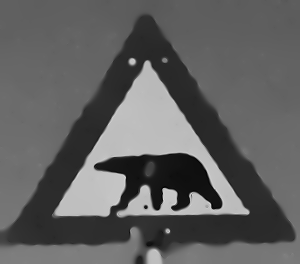} \\[3pt]
\includegraphics[width=0.23\linewidth]
{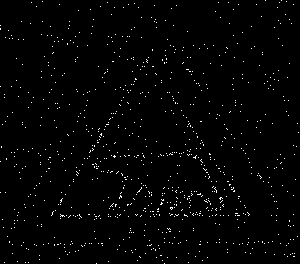} & 
\includegraphics[width=0.23\linewidth]
{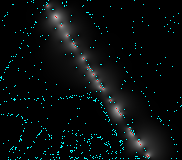} & 
\includegraphics[width=0.23\linewidth]
{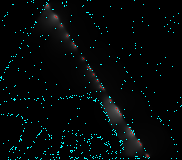} & 
\includegraphics[width=0.23\linewidth]
{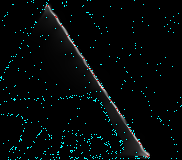} \\
\small{(e) mask} &
\small{(f) cumul. echo} &
\small{(g) cumul. echo} &
\small{(h) cumul. echo}
\end{tabular}
\caption{\label{fig:inp-svalbard}
Inpainting results of the test image \emph{svalbard} using
(b) homogeneous diffusion (HD) inpainting, 
(c) isotropic nonlinear diffusion (NLD) inpainting ($\lambda=0.4$) with 
Charbonnier diffusivity,
(d) EED inpainting ($\lambda=0.3$, $\sigma=1.0$), and zoom into corresponding 
cumulative echoes along a prominent image edge. The mask pixels corresponding 
to echoes are marked in red, and the other mask pixels in cyan.
The three cumulative echoes are rescaled jointly.
We use a mask with $1.5\,\%$ density that is optimised for homogeneous diffusion 
inpainting~\cite{BBBW09}.
The echoes show how EED is able to connect edges even if the mask density
is low and the mask is suboptimal.
}
\end{figure}

Lastly, we consider a more realistic test example in \Cref{fig:inp-svalbard}. 
We use the test image \emph{svalbard} and a mask with $1.5\,\%$ density, which 
is created with a simple version of an optimisation strategy for homogeneous 
diffusion inpainting~\cite{BBBW09}. In this example, the mask pixels are not 
placed perfectly next to the considered edge but might sit a little further 
from it. We see that EED still reconstructs a sharp edge, although it is not 
perfectly straight at locations of low density. Nevertheless, the reconstruction 
is convincing.


\subsubsection{Osmosis Echoes}

\begin{figure}[!tb]
\centering
\tabcolsep3pt
\fboxsep0pt
\begin{tabular}{ccccc}
\raisebox{0.115\linewidth}{\rotatebox[origin=c]{90}{\small{Source Echo}}} &
\includegraphics[width=0.23\linewidth]
{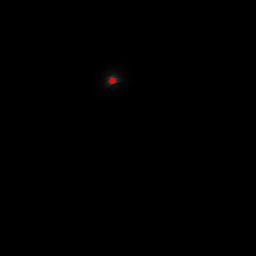} & 
\includegraphics[width=0.23\linewidth]
{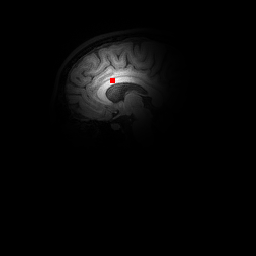} & 
\includegraphics[width=0.23\linewidth]
{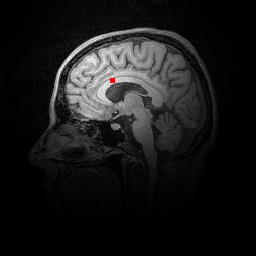} & 
\includegraphics[width=0.23\linewidth]
{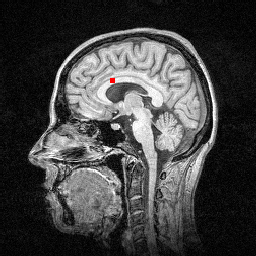} \\[3pt]
\raisebox{0.115\linewidth}{\rotatebox[origin=c]{90}{\small{Drain Echo}}} &
\includegraphics[width=0.23\linewidth]
{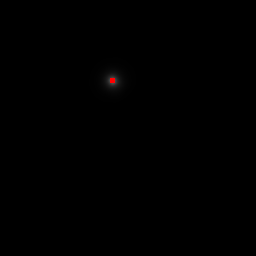} & 
\includegraphics[width=0.23\linewidth]
{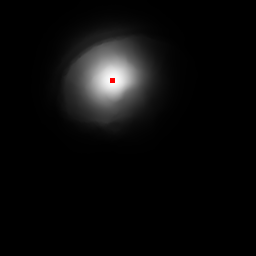} & 
\includegraphics[width=0.23\linewidth]
{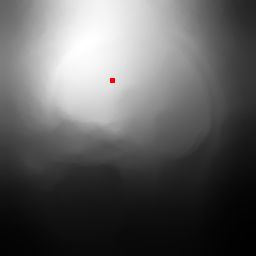} & 
\fbox{\includegraphics[width=0.23\linewidth]
{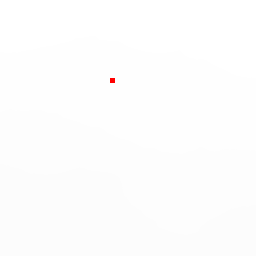}} \\
& \small{(a) $T=25$} & \small{(b) $T=250$} & 
\small{(c) $T=2500$} & \small{(d) $T=500000$}
\end{tabular}
\caption{\label{fig:osm-echoes} Example echoes from osmosis evolution from 
\Cref{fig:osm-evol}. First row: source echo evolution in $(112,80)$. 
Second row: drain echo evolution in $(112,80)$. We see that the source
echo converges to the (rescaled) guidance image, and the drain echo to a
constant image.}
\end{figure}

We have seen in \Cref{subsec:osmosis-echo} that osmosis converges to a 
nonconstant steady state, which is characterised by equal echoes at all 
locations. We illustrate this in \Cref{fig:osm-echoes} by considering the
compatible case. 
We visualise the source and drain echoes at a given location at different 
times throughout the evolutions and show how the echoes converge towards the 
rescaled guidance image and a constant image, just as the theory prescribes.


\subsubsection{Optic Flow Echoes}

Optic flow models are complex, but understanding them as smoothed versions
of the normal flow helps us to get a better intuition. In the following, we
display echoes that show how this regularisation acts.
Similarly to the inpainting case, since
\begin{equation}
\bm{w} = 
    \begin{pmatrix}
        \bm{u} \\ \bm{v}
    \end{pmatrix}
    = \bm{S} 
    \begin{pmatrix}
        \bm{u}_n \\ \bm{v}_v
    \end{pmatrix}
    = \bm{S} \bm{w}_n = \sum_{i=1}^{2N} w_{n, i} \bm{s}_i,
\end{equation}
only the source echoes $\bm{s}_i$ corresponding to locations where the normal 
flow is different from zero influence the result. 
However, in contrast to the inpainting case, there is no mechanism to avoid 
that the flow at these locations can be changed as well.

We visualise optic flow source echoes to see how the information from the 
normal flow is propagated throughout the whole image domain to create a dense 
flow field. To this end, we consider a simplistic test image of a moving 
rectangle.

\begin{figure}[!tb]
\centering
\tabcolsep3pt
\begin{tabular}{cccc}
\small{(a) ref. frame} &
\small{(b) HS flow} & 
\small{(c) zoom into (a)} & 
\small{(d) HS echo zoom} \\[0.5mm]
\includegraphics[width=0.22\linewidth]
{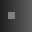} & 
\includegraphics[width=0.22\linewidth]
{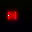} & 
\includegraphics[width=0.22\linewidth]
{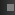} & 
\includegraphics[width=0.22\linewidth]
{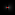} \\[3pt]
\includegraphics[width=0.22\linewidth]
{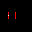} & 
\includegraphics[width=0.22\linewidth]
{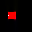} & 
\includegraphics[width=0.22\linewidth]
{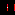} & 
\includegraphics[width=0.22\linewidth]
{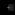} \\
\small{(e) normal flow} &
\small{(f) NE flow} &
\small{(g) zoom into (e)} &
\small{(h) NE echo zoom}
\end{tabular}
\caption{\label{fig:flow-ex-0}
Computed colour-coded optic flow of a square moving one pixel to the right 
and example echo. The flow is computed with the Horn--Schunck (HS) model 
($\alpha=10$) and the Nagel--Enkelmann (NE) method ($\alpha=1$, $\lambda=0.5$).
We use the same colour code as in \Cref{fig:hs-of}, so the ground truth 
motion would be a red square.
We visualise echoes from the central pixel of the left vertical line of the 
normal flow. This location is marked in white in the normal flow and the 
computed flow fields and in red in the corresponding echoes.
Since the normal flow in this pixel only has a horizontal 
direction (see (e)), we only consider the echoes for the horizontal component. 
These echoes have a negligible vertical component, so instead of colour coding
them, we logarithmically rescale the horizontal component and visualise them 
in greyscale. This enhances the visibility of the details. We display zooms
of the echoes which correspond to the zooms in (c) and (g).
The echoes (d) and (h) visualise how the normal flow information is locally 
transported to the inside of the square. However, (d) shows that the 
Horn--Schunck model also propagates flow information to the outside of the 
square, which leads to the blurry result from (b).}
\end{figure}

In \Cref{fig:flow-ex-0} we consider a simple shift by one pixel to the
right and compute the flow field with the Horn--Schunck and Nagel--Enkelmann 
methods. The colour-coded ground truth solution is a red rectangle with sharp 
boundaries, which has the same size as the rectangle in the reference frame 
(first frame). We display the reference image and the sparse normal flow, which 
(aside from the corners) consists only of a horizontal component. 
Considering echoes from a pixel on the left edge, we see that a propagation 
and thus a smoothing of the normal flow takes place. 
For the Horn--Schunck method, the regulariser propagates data to both sides of
the edge of the rectangle, leading to a blurry result. The Nagel--Enkelmann 
method, steered by the discontinuities of the reference image, transports data 
only into the rectangle and thus provides a sharp flow field.


\subsection{Echo Compression}
\label{sec:compression-results}

We now evaluate the compression approach presented in 
\Cref{subsec:compression-approach}.

For inpainting, the number of nonzero echoes is small, and for osmosis 
all echoes are redundant. Furthermore, we can quickly generate echoes
for bilateral filtering and NL means from the explicit weights.
However, for diffusion filters, echoes might vary from pixel to pixel, and a
computation of an echo is costly, since it requires the application of the 
(transposed) state transition matrix, which equates to an entire diffusion 
process. Therefore, we test our approach with some of the diffusion 
filters presented in \Cref{subsubsec:diff-echo}. 

We use the $256 \times 256$ test image \emph{head} with a grey value range 
of $[0, 255]$. The parameters of the methods are selected so that they perform 
a comparable amount of smoothing.
We then compute the probabilistic truncated SVD for each of the filters, 
truncating at $0.5\,\%$, $1.25\,\%$, $2.5\,\%$ and $5\,\%$ of the singular 
values. We select $q=3$ and $\ell=10$. Lastly, we reconstruct the state 
transition matrix with the approximated singular values and vectors. 

\begin{figure}[!tb]
\centering
\tabcolsep3pt
\begin{tabular}{cccc}
\includegraphics[width=0.23\linewidth]
{images/head.png} & 
\includegraphics[width=0.23\linewidth]
{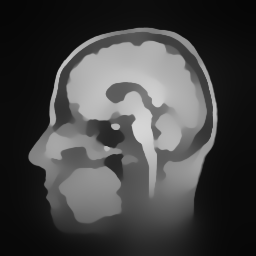} &
\includegraphics[width=0.23\linewidth]
{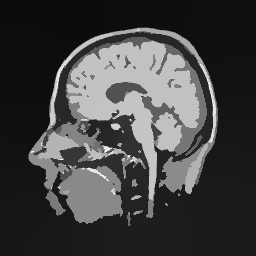} &
\includegraphics[width=0.23\linewidth]
{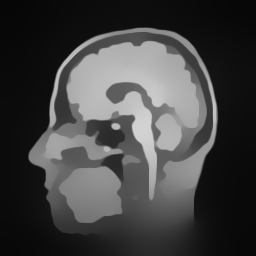} \\
\small{(a) original} &
\small{(b) NLD (PM)} & 
\small{(c) NLD (We)} &
\small{(d) EED (PM)}
\end{tabular}
\caption{\label{fig:head-compression-filtered} 
The test cases for the echo compression method.
(a) The original test image \emph{head}, and filtered results with 
(b) isotropic nolinear diffusion (NLD) with rational Perona--Malik 
diffusivity (PM, $t=190$, $\lambda=3$, $\sigma=0.5$), 
(c) isotropic nolinear diffusion (NLD) with Weickert diffusivity 
(We, $t=15000$, $\lambda=5$, $\sigma=0.5$), and
(d) with edge-enhancing diffusion (EED) with rational Perona--Malik 
diffusivity ($t=280$, $\lambda=3$, $\sigma=0.5$).
The parameters are chosen such that a comparable smoothing effect is 
achieved.
}
\end{figure}

We use the conjugate gradient (CG) method to solve the linear systems in the 
semi-implicit schemes and make use of LAPACK~\cite{ABB+99} to compute the 
singular value and QR decompositions in the RSVD algorithm.
The test image \emph{head} and the filtered versions of three different 
diffusion filters, which we use as test cases, are shown in 
\Cref{fig:head-compression-filtered}.


To evaluate the results, we proceed twofold. First, we compute the 
error $\normin{\bm{S} - \widehat{\bm{S}}}_F$ between the state 
transition matrix and the approximation. Since a calculation of this would 
require us to compute $\bm{S}$ explicitly, we instead apply Hutchinson's trace 
estimator~\cite{Hu90} with Gaussian random vectors~\cite{SR97}, using that
$\normin{\bm{A}}_F^2 = \trace (\bm{A}^T\bm{A})$. 
We select the number of samples based on the results from Avron and 
Toledo~\cite{AT11}. Since we do not need the highest precision, we select the 
number so that the median relative approximation error is at most 10\%. 
However, experiments in~\cite{AT11} suggest that we can expect substantially 
higher accuracy in practice. Nevertheless, we also complement this 
approximation-based quality criterion with a deterministic one: We compute 
the MSE between the filtered image $\bm{u}$ and the reconstruction 
$\widehat{\bm{S}}\bm{f}$ based on the approximated state transition matrix. 
In addition to the quantitative evaluation, we visualise a few reconstructed 
source echoes to get a better intuition about the reconstruction quality.

Then we compare the singular value spectra of the methods. To this end, we
plot the first $5\%$ of the approximated singular values. Since the 
quality of the reconstruction is directly related to the spectrum 
\labelcref{eq:trunc-svd-error}, this gives us an idea of how the quality 
changes w.r.t.\ the compression ratio. Lastly, we visualise some of the 
singular vectors to get a better understanding of how the SVD-based 
representation works.


\begin{table}[!tb]
\caption{Errors for RSVD-based compression. Estimated~\cite{Hu90} Frobenius 
norm of the error and MSE of the reconstruction of the filtered image 
depending on the fraction of used singular values.}
\label{tab:rsvd-600}
\centering
\tabcolsep5pt
\begin{tabular}{l|c|c|c|c}
  percentage of & \multicolumn{4}{|c}{Frobenius Norm / MSE}\\
  singular values & $0.5\,\%$ & $1.25\,\%$ & $2.5\,\%$ & $5.0\,\%$\\\hline
  NLD (PM) & $2.198$ / $0.84810$ & $0.666$ / $0.02864$ 
  & $0.074$ / $0.00034$ & $0.012$ / $0.00010$ \\
  NLD (We) & $37.55$ / $6451.53$ & $30.95$ / $3637.93$ 
  & $16.28$ / $534.286$ & $0.008$ / $0.00291$ \\
  EED (PM) & $0.015$ / $0.00012$ & $0.012$ / $0.00010$ 
  & $0.012$ / $0.00010$ & $0.012$ / $0.00010$
\end{tabular}
\end{table}


\begin{figure}[!tb]
\centering
\tabcolsep2pt
\begin{tabular}{ccccc|c}
\raisebox{0.09\linewidth}{\rotatebox[origin=c]{90}{\scriptsize{NLD (PM)}}} &
\includegraphics[width=0.18\linewidth]
{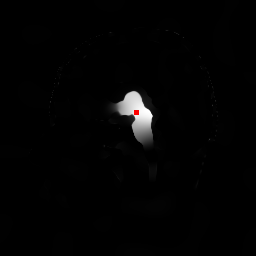} & 
\includegraphics[width=0.18\linewidth]
{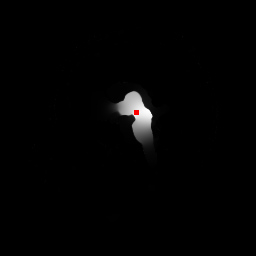} & 
\includegraphics[width=0.18\linewidth]
{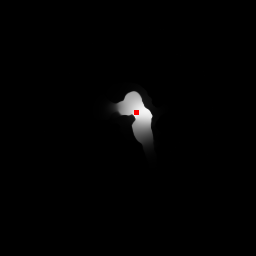} & 
\includegraphics[width=0.18\linewidth]
{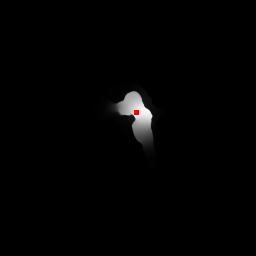} & 
\includegraphics[width=0.18\linewidth]
{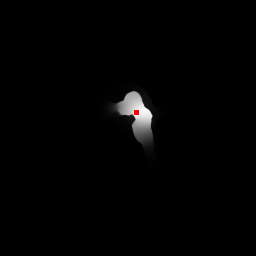} \\
\raisebox{0.09\linewidth}{\rotatebox[origin=c]{90}{\scriptsize{NLD (We)}}} &
\includegraphics[width=0.18\linewidth]
{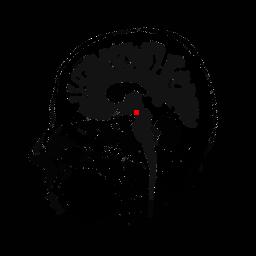} & 
\includegraphics[width=0.18\linewidth]
{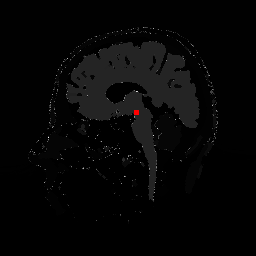} & 
\includegraphics[width=0.18\linewidth]
{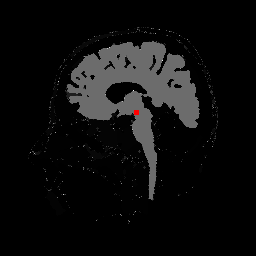} & 
\includegraphics[width=0.18\linewidth]
{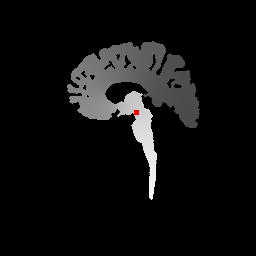} & 
\includegraphics[width=0.18\linewidth]
{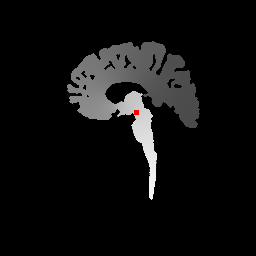} \\
&
(a) $0.5\%$ &
(b) $1.25\%$ &
(c) $2.5\%$ & 
(d) $5.0\%$ &
(e) original \\[0.5mm]
\end{tabular}
\caption{\label{fig:head-echoes-reco} 
Reconstructed source echoes for nonlinear diffusion at location $(112, 136)$, 
using different percentages of the singular values and the original echo
for comparison. The red dot marks the echo location.
We see that $2.5\,\%$ of singular values suffice for a visually error-free 
reconstruction for the Perona--Malik diffusivity, while the reconstruction
with $1.25\,\%$ has almost negligible brightness differences. 
For the Weickert diffusivity, $5\,\%$ of the singular values are required to 
adequately capture all details.}
\end{figure}


We see in \Cref{tab:rsvd-600} that the echoes corresponding to EED have the 
best compression potential, with isotropic nolinear diffusion with the 
rational Perona--Malik diffusivity slightly behind. 
The results with nonlinear diffusion and the \mbox{Weickert} diffusivity stand
in contrast to that. We have seen in the segmentation experiments in 
\Cref{subsubsec:exp-segmentation} that the echoes extend over full 
segments, which at first glance would imply a lot of redundancy and great
compression potential. However, the results show that the echoes require
the most data to be adequately reconstructed. The reason for this lies in a 
specific phenomenon: Image edges are typically not perfectly sharp step edges, 
but may extend over a larger distance, affecting more pixels. If the edge is 
strong, the contrast between the individual pixels might still be high. 
This means that the inner edge pixels do not experience any smoothing at all, 
thus their corresponding echoes are localised almost entirely within the 
position. These impulse-like echoes are not easily encoded in the SVD basis 
and increase the number of large singular values, deteriorating the entire 
compression quality.

In \Cref{fig:head-echoes-reco} we visually compare reconstructions for the
two isotropic nonlinear diffusion models. We see that the reconstruction for 
the rational Perona--Malik diffusivity which uses $2.5\,\%$ of the singular 
values is visually indistinguishable from the original, while the 
reconstructions using $0.5\,\%$ and $1.25\,\%$ have minor brightness 
deviations. This gives us a feeling for the values of the error norm that we 
should target. For the Weickert diffusivity, only the reconstruction with 
$5\,\%$ is adequate.

\begin{figure}[!tb]
\centering
\includegraphics[width=0.9\linewidth]{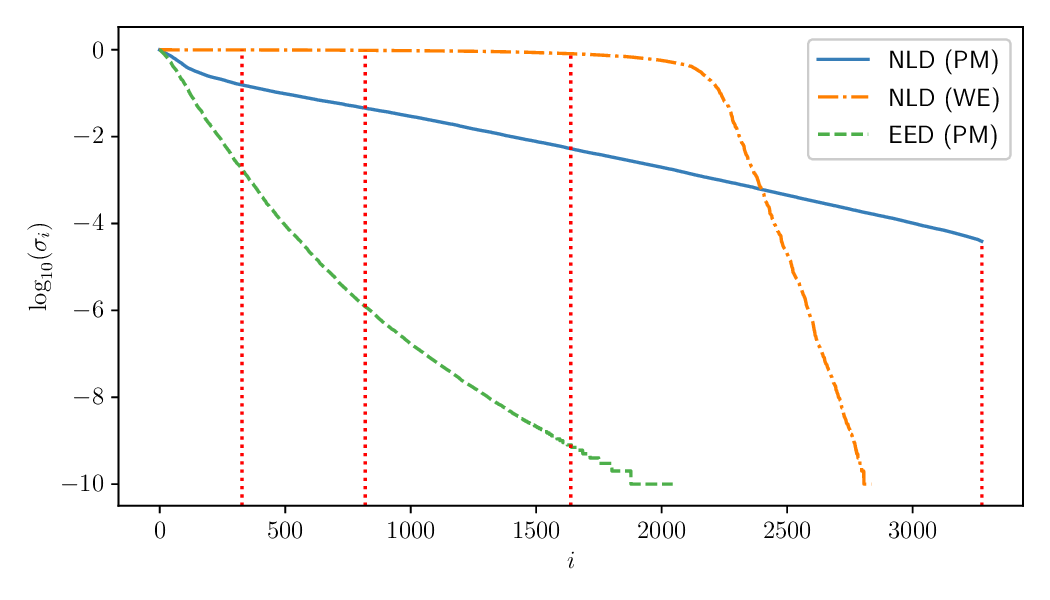}
\caption{\label{fig:spectra} The singular value spectra from the experiments.
The vertical lines depict $0.5\,\%$, $1.25\,\%$, $2.5\,\%$ and $5.0\,\%$ of 
the singular values. Note that the Weickert diffusivity leads to a lot of 
similarly large singular values and a fast drop, while the two methods with 
the rational Perona--Malik diffusivity have a more even decay.}
\end{figure}

Let us now look at the spectra in \Cref{fig:spectra}. They not 
only confirm the results in \Cref{tab:rsvd-600}, but also provide additional 
insights. It is clear to see that the Weickert diffusivity leads to a large 
number of singular values, before we get a sudden decay.
The large singular values correspond to the individual segments and pointwise
echoes that barely overlap, so they either can or cannot be represented 
adequately. Isotropic nonlinear diffusion with the rational Perona--Malik 
diffusivity leads to echoes with larger overlap, since the edge preservation 
mechanism is less prohibitive, and pointwise echoes are not an issue. This is 
reflected in the spectrum, which has a more even decay. For EED we see that 
the singular values fall off even faster. Due to the additional smoothing along 
the edges, the echoes extend over a larger area, creating more redundancy. 
This enables a very efficient compression.


\begin{figure}[!htb]
\centering
\tabcolsep2pt
\fboxsep=0pt
\begin{tabular}{cccccc}
\raisebox{0.11\linewidth}{\rotatebox[origin=c]{90}{\scriptsize{$k=1$}}} &
\includegraphics[width=0.22\linewidth]
{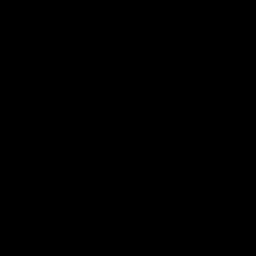} & 
\fbox{\includegraphics[width=0.22\linewidth]
{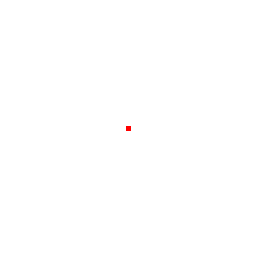}} &
\raisebox{0.11\linewidth}{\rotatebox[origin=c]{90}{\scriptsize{$k=10$}}} &
\includegraphics[width=0.22\linewidth]
{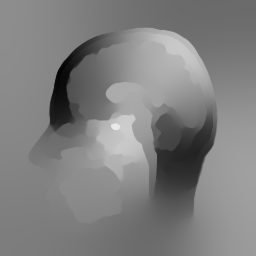} & 
\includegraphics[width=0.22\linewidth]
{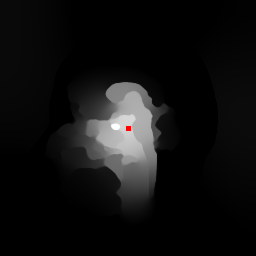} \\[2pt]
\raisebox{0.11\linewidth}{\rotatebox[origin=c]{90}{\scriptsize{$k=2$}}} &
\includegraphics[width=0.22\linewidth]
{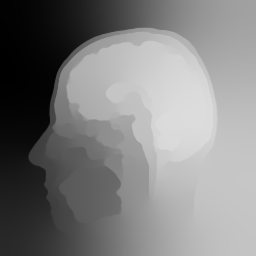} & 
\includegraphics[width=0.22\linewidth]
{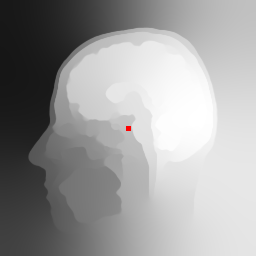} & 
\raisebox{0.11\linewidth}{\rotatebox[origin=c]{90}{\scriptsize{$k=25$}}} &
\includegraphics[width=0.22\linewidth]
{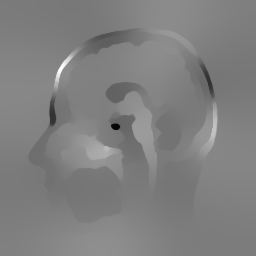} & 
\includegraphics[width=0.22\linewidth]
{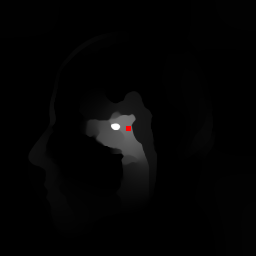}  \\[2pt]
\raisebox{0.11\linewidth}{\rotatebox[origin=c]{90}{\scriptsize{$k=3$}}} &
\includegraphics[width=0.22\linewidth]
{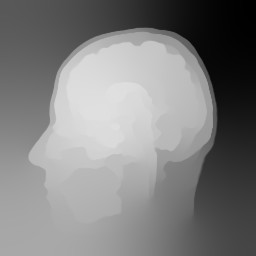} & 
\includegraphics[width=0.22\linewidth]
{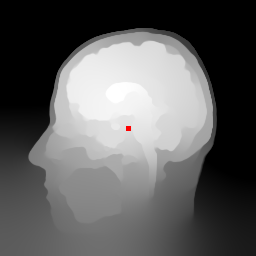} & 
\raisebox{0.11\linewidth}{\rotatebox[origin=c]{90}{\scriptsize{$k=50$}}} &
\includegraphics[width=0.22\linewidth]
{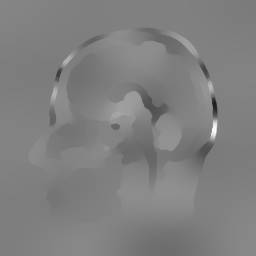} & 
\includegraphics[width=0.22\linewidth]
{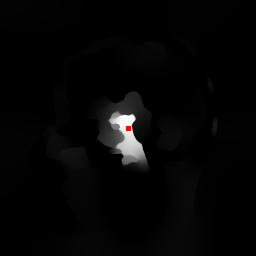} \\[2pt]
\raisebox{0.11\linewidth}{\rotatebox[origin=c]{90}{\scriptsize{$k=4$}}} &
\includegraphics[width=0.22\linewidth]
{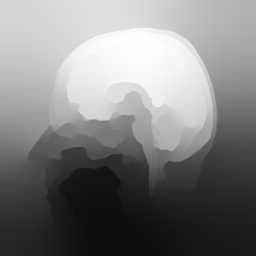} & 
\includegraphics[width=0.22\linewidth]
{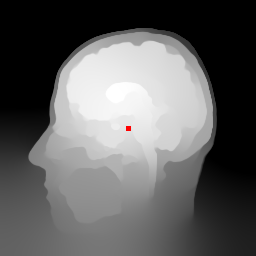} & 
\raisebox{0.11\linewidth}{\rotatebox[origin=c]{90}{\scriptsize{$k=100$}}} &
\includegraphics[width=0.22\linewidth]
{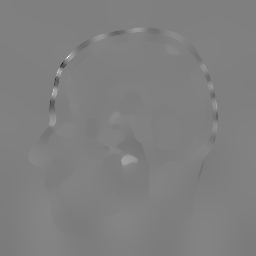}  & 
\includegraphics[width=0.22\linewidth]
{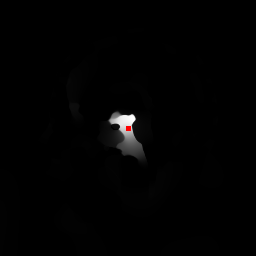} \\[2pt]
\raisebox{0.11\linewidth}{\rotatebox[origin=c]{90}{\scriptsize{$k=5$}}} &
\includegraphics[width=0.22\linewidth]
{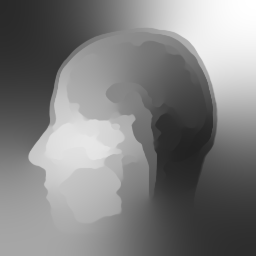} & 
\includegraphics[width=0.22\linewidth]
{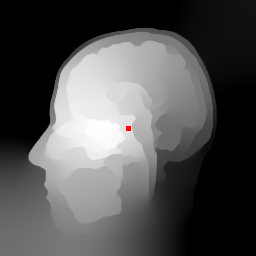} &
\raisebox{0.11\linewidth}{\rotatebox[origin=c]{90}{\scriptsize{$k=1000$}}} &
\includegraphics[width=0.22\linewidth]
{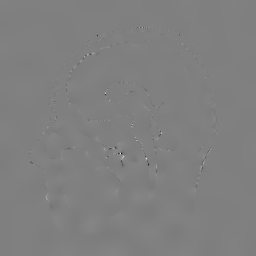} & 
\includegraphics[width=0.22\linewidth]
{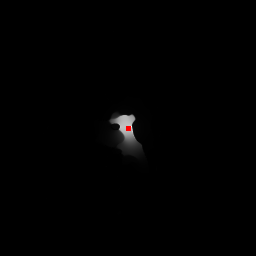} \\[2pt]
\raisebox{0.11\linewidth}{\rotatebox[origin=c]{90}{\scriptsize{$k=7$}}} &
\includegraphics[width=0.22\linewidth]
{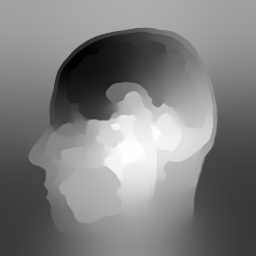} & 
\includegraphics[width=0.22\linewidth]
{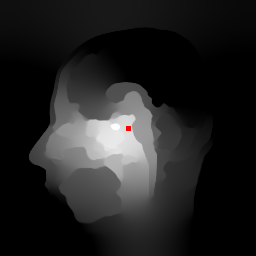} &
\raisebox{0.11\linewidth}{\rotatebox[origin=c]{90}{\scriptsize{original}}} 
& & \includegraphics[width=0.22\linewidth]
{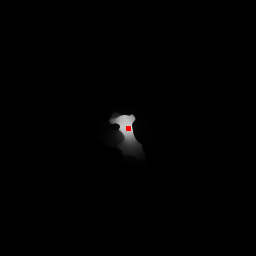} \\
& \small{$[\bm{U}]_k$} & \small{echo reconstruction} &
& \small{$[\bm{U}]_k$} & \small{echo reconstruction}
\end{tabular}
\caption{\label{fig:svd-inspection} 
Plot of some of the left singular vectors $[\bm{U}]_k$ and the corresponding 
reconstructions of the source echo at $(128,128)$. We use the test 
case with nonlinear diffusion with the rational Perona--Malik diffusivity 
from the previous experiments.
Note how the singular vectors become more and more detailed, allowing for
an accurate reconstruction of small structures when using more data.}
\end{figure}

As a last step, plotting some of the singular vectors and the corresponding 
reconstructions using all the preceding SVD information shows how adding more 
SVD components refines the echo reconstruction step by step. As an example, 
we select isotropic nonlinear diffusion with the rational Perona--Malik 
diffusivity. In \Cref{fig:svd-inspection} we plot some of the left singular 
vectors as well as the reconstructions of the source echo at a central pixel, 
using all the SVD data up to the respective index. Since singular vectors also 
have negative values, we shift the value of $0$ to $127.5$.


The first singular vector corresponds to a flat image, since the process 
converges to the average grey value of the given image.
The next singular vectors are an overlay of the most important structures
of the image, where the structures become more and more detailed and 
smaller in their spatial extent. The reconstruction of the echo converges
to the original by discarding more and more irrelevant information, which is
possible due to the fine details contained in later singular vectors.
We see that the $100$-th singular vector contains small-scale details, which 
is why the reconstruction is already fairly accurate.
The reconstruction using $1000$ singular vectors is almost indistinguishable
from the original, which is in line with the results from \Cref{tab:rsvd-600}
and \Cref{fig:spectra}.


\subsubsection{An Extension for Rapidly Decaying Diffusivities}

We have seen in the results in \Cref{tab:rsvd-600} and the spectrum in 
\Cref{fig:spectra} that the Weickert diffusivity suffers from the pointwise 
echoes. To mitigate this, we propose an exclusion mechanism.

To this end, we detect echoes for which the central pixel (i.e.~the 
corresponding diagonal element of the state transition matrix $\bm{S}$) is 
larger than some threshold $1 - \epsilon$. We simply store the corresponding 
echo location (i.e.~the two pixel coordinates) and exclude the corresponding 
row and column from the state transition matrix, leading to a smaller matrix, 
which we compress with the method described in 
\Cref{subsec:compression-approach}.
At decompression, we simply add the discarded echoes as unit impulses.
The parameter $\epsilon$ leads to a trade-off between the number of echoes
that we exclude (\emph{efficiency}) and the error that we make by describing 
the corresponding echo as a unit impulse (\emph{accuracy}).

To find the pixels that should be excluded, we first need to compute the
corresponding echoes, increasing the cost of the encoding phase. 
However, we can reduce this additional cost by decreasing the number of
potential exclusion candidates for which the echoes must be computed.
To this end, recall that the state transition matrix $\bm{S}$ for
diffusion processes has nonnegative entries and unit row sums  (see 
\Cref{subsubsec:diff-echo}), and that the initial image $\bm{f}$ is also 
nonnegative. Then it follows from \labelcref{eq:drain-echo-reco} that if the 
diagonal element $s_{ii} > 1-\epsilon$, we must have 
$(1-\epsilon) \, f_i + \epsilon \, f_{min} 
< u_{i} < (1-\epsilon) \, f_i + \epsilon \, f_{max}$
for the corresponding pixel $u_i$ of the filtered image.
If any of these inequalities is violated, we can conclude that 
$s_{ii} \leq 1-\epsilon$ and the corresponding echo will not be excluded. 
However, fulfilling the inequalities does not guarantee  
$s_{ii} > 1-\epsilon$, so we need to manually check by explicitly computing 
the echo.

We test our proposed approach by choosing $\epsilon \in \{0.0, 0.05, 0.1\}$ and 
compressing the echoes from isotropic nonlinear diffusion with the Weickert 
diffusivity and the parameters from before.


\begin{table}[!tb]
\caption{Errors for isotropic nonlinear diffusion with the Weickert diffusivity
using the proposed exclusion mechanism with $\epsilon \in \{0.0, 0.05, 0.1\}$.
Estimated~\cite{Hu90} Frobenius norm of the error and MSE of the reconstruction 
of the filtered image depending on the fraction of used singular values.}
\label{tab:we-excl}
\centering
\tabcolsep5pt
\begin{tabular}{l|c|c|c|c}
  percentage of & \multicolumn{4}{|c}{Frobenius Norm / MSE}\\
  singular values & $0.5\,\%$ & $1.25\,\%$ & $2.5\,\%$ & $5.0\,\%$\\\hline
  $\epsilon=0.0$ & $37.55$ / $6451.53$ & $30.95$ / $3637.93$ 
  & $16.28$ / $534.286$ & $\bm{0.008}$ / $\bm{0.00291}$ \\
  $\epsilon=0.05$ & $22.38$ / $2736.83$  & $12.17$ / $269.923$ 
  & $\bm{0.745}$ / $\bm{0.01346}$ & $0.745$ / $0.01358$ \\
  $\epsilon=0.1$ & $\bm{16.62}$ / $\bm{1282.20}$ & $\bm{4.528}$ / $\bm{9.19979}$ 
  & $1.596$ / $0.05048$ & $1.596$ / $0.05048$
\end{tabular}
\end{table}


\begin{figure}[!tb]
\centering
\includegraphics[width=0.9\linewidth]{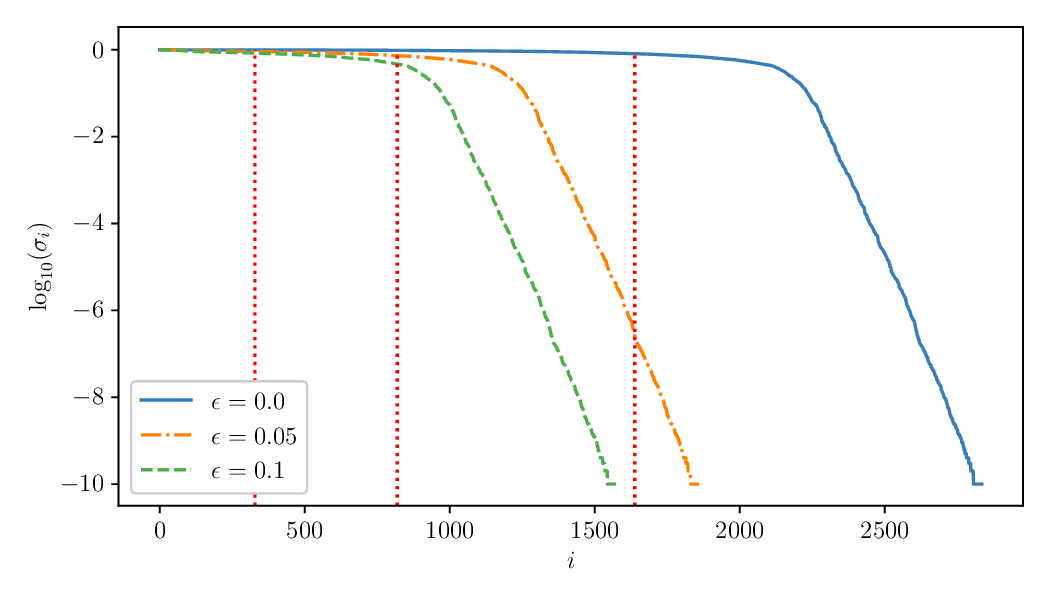}
\caption{\label{fig:spectra-excl} The singular value spectra using our 
exclusion approach with different values for $\epsilon$.
The vertical lines depict $0.5\,\%$, $1.25\,\%$ and $2.5\,\%$ of 
the singular values. We see that the proposed exclusion mechanism shifts the
spectrum and decreases the number of large singular values. Note that the
spectrum does not reflect the approximation error that is made by describing
an echo as a unit impulse.}
\end{figure}


The results in \Cref{tab:we-excl} and \Cref{fig:spectra-excl} confirm the 
effectiveness of our proposed exclusion mechanism. \Cref{tab:we-excl} shows 
that we can get good results at high compression ratios, since few singular 
vectors now suffice to capture the most important structures of the echoes. 
This is confirmed by the singular value spectra in \Cref{fig:spectra-excl}, 
which show a shift in the curve and a decay that occurs much earlier.
However, one must keep in mind that this approach introduces a lower bound
on the reconstruction error, which becomes apparent for the results at 
$5\,\%$ of the singular values. Therefore, the parameter $\epsilon$ should 
be adjusted to the requirements.

Although the required precomputation of the echoes for exclusion  
increases the cost of the encoding phase, it should be further noted that 
the exclusion approach even reduces the storage cost slightly, since each 
excluded point reduces the size of the singular vectors while only requiring 
the storage of two integers.


\section{Conclusions}
\label{sec:conclusions}

We have introduced the filter echo as a general visualisation framework
that can be applied to a large number of filters from image processing and
even computer vision.
While it includes filters for which a visualisation of the kernel  
has been done previously, we have shown that it can be readily extended to 
other filters, such as osmosis or inpainting, and even to optic flow models.

We have presented its capabilities of visualising the inner workings of
complex filters, which enables a better understanding, comparability between
filters, and can even be helpful for segmentation.

In addition, in the present paper and its conference predecessor~\cite{GWFC25}, 
we have proposed the first compression approach specifically
tailored towards filter echoes, which counteracts the extensive storage 
cost of the filter echo. We have shown on a number of test cases with
diffusion models that an SVD-based approach can drastically reduce the storage 
requirements while preserving visual quality. Furthermore, we have seen
that reconstruction is simple and can be achieved by a single matrix-vector
multiplication. This compression approach completes the filter echo framework 
and makes it relevant for practical applications where a large number of 
echoes must be readily available.

In our future work, our aim is to compute filter echoes for further filters 
for which the interpretation of the echo is not straightforward.


\backmatter

\section*{Declarations}

\bmhead{Disclosure of Interests} 
The authors have no competing interests to declare that are relevant to the 
content of this article.
\bmhead{Data/Code Availability} 
The datasets used and/or analysed during the current study are available from 
the corresponding author on reasonable request.
\bmhead{Authors' Contributions} 
DG and JW both developed concepts and provided theory. DG took care of all 
implementations, partially based on code by JW, and performed all experiments. 
DG wrote the paper with feedback and modifications from JW. 
IF and \"{O}\c{C} performed earlier research in their student theses, which 
gave the inspiration for some parts of the conducted work. 
All authors have read and approved the final manuscript.


\bibliography{echo-references}


\begin{thebibliography}{66}
\ifx \bisbn   \undefined \def \bisbn  #1{ISBN #1}\fi
\ifx \binits  \undefined \def \binits#1{#1}\fi
\ifx \bauthor  \undefined \def \bauthor#1{#1}\fi
\ifx \batitle  \undefined \def \batitle#1{#1}\fi
\ifx \bjtitle  \undefined \def \bjtitle#1{#1}\fi
\ifx \bvolume  \undefined \def \bvolume#1{\textbf{#1}}\fi
\ifx \byear  \undefined \def \byear#1{#1}\fi
\ifx \bissue  \undefined \def \bissue#1{#1}\fi
\ifx \bfpage  \undefined \def \bfpage#1{#1}\fi
\ifx \blpage  \undefined \def \blpage #1{#1}\fi
\ifx \burl  \undefined \def \burl#1{\textsf{#1}}\fi
\ifx \doiurl  \undefined \def \doiurl#1{\url{https://doi.org/#1}}\fi
\ifx \betal  \undefined \def \betal{\textit{et al.}}\fi
\ifx \binstitute  \undefined \def \binstitute#1{#1}\fi
\ifx \binstitutionaled  \undefined \def \binstitutionaled#1{#1}\fi
\ifx \bctitle  \undefined \def \bctitle#1{#1}\fi
\ifx \beditor  \undefined \def \beditor#1{#1}\fi
\ifx \bpublisher  \undefined \def \bpublisher#1{#1}\fi
\ifx \bbtitle  \undefined \def \bbtitle#1{#1}\fi
\ifx \bedition  \undefined \def \bedition#1{#1}\fi
\ifx \bseriesno  \undefined \def \bseriesno#1{#1}\fi
\ifx \blocation  \undefined \def \blocation#1{#1}\fi
\ifx \bsertitle  \undefined \def \bsertitle#1{#1}\fi
\ifx \bsnm \undefined \def \bsnm#1{#1}\fi
\ifx \bsuffix \undefined \def \bsuffix#1{#1}\fi
\ifx \bparticle \undefined \def \bparticle#1{#1}\fi
\ifx \barticle \undefined \def \barticle#1{#1}\fi
\bibcommenthead
\ifx \bconfdate \undefined \def \bconfdate #1{#1}\fi
\ifx \botherref \undefined \def \botherref #1{#1}\fi
\ifx \url \undefined \def \url#1{\textsf{#1}}\fi
\ifx \bchapter \undefined \def \bchapter#1{#1}\fi
\ifx \bbook \undefined \def \bbook#1{#1}\fi
\ifx \bcomment \undefined \def \bcomment#1{#1}\fi
\ifx \oauthor \undefined \def \oauthor#1{#1}\fi
\ifx \citeauthoryear \undefined \def \citeauthoryear#1{#1}\fi
\ifx \endbibitem  \undefined \def \endbibitem {}\fi
\ifx \bconflocation  \undefined \def \bconflocation#1{#1}\fi
\ifx \arxivurl  \undefined \def \arxivurl#1{\textsf{#1}}\fi
\csname PreBibitemsHook\endcsname

\bibitem[\protect\citeauthoryear{Aurich and Weule}{1995}]{AW95}
\begin{bchapter}
\bauthor{\bsnm{Aurich}, \binits{V.}},
\bauthor{\bsnm{Weule}, \binits{J.}}:
\bctitle{Non-linear {G}aussian filters performing edge preserving diffusion}.
In: \beditor{\bsnm{Sagerer}, \binits{G.}},
\beditor{\bsnm{Posch}, \binits{S.}},
\beditor{\bsnm{Kummert}, \binits{F.}} (eds.)
\bbtitle{Mustererkennung 1995},
pp. \bfpage{538}--\blpage{545}.
\bpublisher{Springer},
\blocation{Berlin}
(\byear{1995})
\end{bchapter}
\endbibitem

\bibitem[\protect\citeauthoryear{Smith and Brady}{1997}]{SB97}
\begin{barticle}
\bauthor{\bsnm{Smith}, \binits{S.M.}},
\bauthor{\bsnm{Brady}, \binits{J.M.}}:
\batitle{{SUSAN}: A new approach to low-level image processing}.
\bjtitle{International Journal of Computer Vision}
\bvolume{23}(\bissue{1}),
\bfpage{45}--\blpage{78}
(\byear{1997})
\end{barticle}
\endbibitem

\bibitem[\protect\citeauthoryear{Tomasi and Manduchi}{1998}]{TM98}
\begin{bchapter}
\bauthor{\bsnm{Tomasi}, \binits{C.}},
\bauthor{\bsnm{Manduchi}, \binits{R.}}:
\bctitle{Bilateral filtering for gray and color images}.
In: \bbtitle{Proc. Sixth International Conference on Computer Vision},
pp. \bfpage{839}--\blpage{846}.
\bpublisher{Narosa Publishing House},
\blocation{Bombay, India}
(\byear{1998})
\end{bchapter}
\endbibitem

\bibitem[\protect\citeauthoryear{Buades et~al.}{2005}]{BCM05a}
\begin{barticle}
\bauthor{\bsnm{Buades}, \binits{A.}},
\bauthor{\bsnm{Coll}, \binits{B.}},
\bauthor{\bsnm{Morel}, \binits{J.-M.}}:
\batitle{A review of image denoising algorithms, with a new one}.
\bjtitle{Multiscale Modeling and Simulation}
\bvolume{4}(\bissue{2}),
\bfpage{490}--\blpage{530}
(\byear{2005})
\end{barticle}
\endbibitem

\bibitem[\protect\citeauthoryear{Perona and Malik}{1990}]{PM90}
\begin{barticle}
\bauthor{\bsnm{Perona}, \binits{P.}},
\bauthor{\bsnm{Malik}, \binits{J.}}:
\batitle{Scale space and edge detection using anisotropic diffusion}.
\bjtitle{IEEE Transactions on Pattern Analysis and Machine Intelligence}
\bvolume{12},
\bfpage{629}--\blpage{639}
(\byear{1990})
\end{barticle}
\endbibitem

\bibitem[\protect\citeauthoryear{{ter Haar Romeny}}{1994}]{Ha94}
\begin{bbook}
\beditor{\bsnm{{ter Haar Romeny}}, \binits{B.M.}} (ed.):
\bbtitle{Geometry-Driven Diffusion in Computer Vision}.
\bsertitle{Computational Imaging and Vision},
vol. \bseriesno{1}.
\bpublisher{Kluwer},
\blocation{Dordrecht}
(\byear{1994})
\end{bbook}
\endbibitem

\bibitem[\protect\citeauthoryear{Weickert}{1998}]{We97}
\begin{bbook}
\bauthor{\bsnm{Weickert}, \binits{J.}}:
\bbtitle{Anisotropic Diffusion in Image Processing}.
\bpublisher{Teubner},
\blocation{Stuttgart}
(\byear{1998})
\end{bbook}
\endbibitem

\bibitem[\protect\citeauthoryear{Dam and Nielsen}{2001}]{DN01}
\begin{bchapter}
\bauthor{\bsnm{Dam}, \binits{E.}},
\bauthor{\bsnm{Nielsen}, \binits{M.}}:
\bctitle{Exploring non-linear diffusion: The diffusion echo}.
In: \beditor{\bsnm{Kerckhove}, \binits{M.}} (ed.)
\bbtitle{Scale-Space and Morphology in Computer Vision}.
\bsertitle{Lecture Notes in Computer Science},
vol. \bseriesno{2106},
pp. \bfpage{264}--\blpage{272}.
\bpublisher{Springer},
\blocation{Berlin}
(\byear{2001})
\end{bchapter}
\endbibitem

\bibitem[\protect\citeauthoryear{Proakis and Manolakis}{2013}]{PM13}
\begin{bbook}
\bauthor{\bsnm{Proakis}, \binits{J.}},
\bauthor{\bsnm{Manolakis}, \binits{D.}}:
\bbtitle{Digital Signal Processing},
\bedition{4th} edn.
\bpublisher{Pearson International},
\blocation{London}
(\byear{2013})
\end{bbook}
\endbibitem

\bibitem[\protect\citeauthoryear{Iijima}{1962}]{Ii62}
\begin{barticle}
\bauthor{\bsnm{Iijima}, \binits{T.}}:
\batitle{Basic theory on normalization of pattern (in case of typical
  one-dimensional pattern)}.
\bjtitle{Bulletin of the Electrotechnical Laboratory}
\bvolume{26},
\bfpage{368}--\blpage{388}
(\byear{1962}).
\bcomment{In Japanese}
\end{barticle}
\endbibitem

\bibitem[\protect\citeauthoryear{Aubert and Kornprobst}{2006}]{AK06}
\begin{bbook}
\bauthor{\bsnm{Aubert}, \binits{G.}},
\bauthor{\bsnm{Kornprobst}, \binits{P.}}:
\bbtitle{Mathematical Problems in Image Processing: Partial Differential
  Equations and the Calculus of Variations},
\bedition{2}nd edn.
\bsertitle{Applied Mathematical Sciences},
vol. \bseriesno{147}.
\bpublisher{Springer},
\blocation{New York}
(\byear{2006})
\end{bbook}
\endbibitem

\bibitem[\protect\citeauthoryear{Gali\'c et~al.}{2008}]{GWWB08}
\begin{barticle}
\bauthor{\bsnm{Gali\'c}, \binits{I.}},
\bauthor{\bsnm{Weickert}, \binits{J.}},
\bauthor{\bsnm{Welk}, \binits{M.}},
\bauthor{\bsnm{Bruhn}, \binits{A.}},
\bauthor{\bsnm{Belyaev}, \binits{A.}},
\bauthor{\bsnm{Seidel}, \binits{H.-P.}}:
\batitle{Image compression with anisotropic diffusion}.
\bjtitle{Journal of Mathematical Imaging and Vision}
\bvolume{31}(\bissue{2--3}),
\bfpage{255}--\blpage{269}
(\byear{2008})
\end{barticle}
\endbibitem

\bibitem[\protect\citeauthoryear{Horn and Schunck}{1981}]{HS81}
\begin{barticle}
\bauthor{\bsnm{Horn}, \binits{B.}},
\bauthor{\bsnm{Schunck}, \binits{B.}}:
\batitle{Determining optical flow}.
\bjtitle{Artificial Intelligence}
\bvolume{17},
\bfpage{185}--\blpage{203}
(\byear{1981})
\end{barticle}
\endbibitem

\bibitem[\protect\citeauthoryear{Ikeuchi and Horn}{1981}]{IH81}
\begin{barticle}
\bauthor{\bsnm{Ikeuchi}, \binits{K.}},
\bauthor{\bsnm{Horn}, \binits{B.K.P.}}:
\batitle{Numerical shape from shading and occluding boundaries}.
\bjtitle{Artificial Intelligence}
\bvolume{17}(\bissue{1}),
\bfpage{141}--\blpage{184}
(\byear{1981})
\end{barticle}
\endbibitem

\bibitem[\protect\citeauthoryear{Welk et~al.}{2005}]{WTBW05}
\begin{bchapter}
\bauthor{\bsnm{Welk}, \binits{M.}},
\bauthor{\bsnm{Theis}, \binits{D.}},
\bauthor{\bsnm{Brox}, \binits{T.}},
\bauthor{\bsnm{Weickert}, \binits{J.}}:
\bctitle{{PDE}-based deconvolution with forward-backward diffusivities and
  diffusion tensors}.
In: \beditor{\bsnm{Kimmel}, \binits{R.}},
\beditor{\bsnm{Sochen}, \binits{N.}},
\beditor{\bsnm{Weickert}, \binits{J.}} (eds.)
\bbtitle{Scale Space and {PDE} Methods in Computer Vision}.
\bsertitle{Lecture Notes in Computer Science},
vol. \bseriesno{3459},
pp. \bfpage{585}--\blpage{597}.
\bpublisher{Springer},
\blocation{Berlin}
(\byear{2005})
\end{bchapter}
\endbibitem

\bibitem[\protect\citeauthoryear{Carlsson}{1988}]{Ca88}
\begin{barticle}
\bauthor{\bsnm{Carlsson}, \binits{S.}}:
\batitle{Sketch based coding of grey level images}.
\bjtitle{Signal Processing}
\bvolume{15}(\bissue{1}),
\bfpage{57}--\blpage{83}
(\byear{1988})
\end{barticle}
\endbibitem

\bibitem[\protect\citeauthoryear{Mainberger et~al.}{2011}]{MHWT+11}
\begin{bchapter}
\bauthor{\bsnm{Mainberger}, \binits{M.}},
\bauthor{\bsnm{Hoffmann}, \binits{S.}},
\bauthor{\bsnm{Weickert}, \binits{J.}},
\bauthor{\bsnm{Tang}, \binits{C.H.}},
\bauthor{\bsnm{Johannsen}, \binits{D.}},
\bauthor{\bsnm{Neumann}, \binits{F.}},
\bauthor{\bsnm{Doerr}, \binits{B.}}:
\bctitle{Optimising spatial and tonal data for homogeneous diffusion
  inpainting.}
In: \beditor{\bsnm{Bruckstein}, \binits{A.}},
\beditor{\bsnm{Haar~Romeny}, \binits{B.}},
\beditor{\bsnm{Bronstein}, \binits{A.}},
\beditor{\bsnm{Bronstein}, \binits{M.}} (eds.)
\bbtitle{Scale Space and Variational Methods in Computer Vision}.
\bsertitle{Lecture Notes in Computer Science},
vol. \bseriesno{6667},
pp. \bfpage{26}--\blpage{37}.
\bpublisher{Springer},
\blocation{Berlin}
(\byear{2011})
\end{bchapter}
\endbibitem

\bibitem[\protect\citeauthoryear{Weickert et~al.}{2013}]{WHBV13}
\begin{bchapter}
\bauthor{\bsnm{Weickert}, \binits{J.}},
\bauthor{\bsnm{Hagenburg}, \binits{K.}},
\bauthor{\bsnm{Breu{\ss}}, \binits{M.}},
\bauthor{\bsnm{Vogel}, \binits{O.}}:
\bctitle{Linear osmosis models for visual computing}.
In: \beditor{\bsnm{Heyden}, \binits{A.}},
\beditor{\bsnm{Kahl}, \binits{F.}},
\beditor{\bsnm{Olsson}, \binits{C.}},
\beditor{\bsnm{Oskarsson}, \binits{M.}},
\beditor{\bsnm{Tai}, \binits{X.-C.}} (eds.)
\bbtitle{Energy Minimisation Methods in Computer Vision and Pattern
  Recognition}.
\bsertitle{Lecture Notes in Computer Science},
vol. \bseriesno{8081},
pp. \bfpage{26}--\blpage{39}.
\bpublisher{Springer},
\blocation{Berlin}
(\byear{2013})
\end{bchapter}
\endbibitem

\bibitem[\protect\citeauthoryear{Milanfar}{2013}]{Mi13a}
\begin{barticle}
\bauthor{\bsnm{Milanfar}, \binits{P.}}:
\batitle{A tour of modern image filtering: New insights and methods, both
  practical and theoretical}.
\bjtitle{IEEE Signal Processing Magazine}
\bvolume{30}(\bissue{1}),
\bfpage{106}--\blpage{128}
(\byear{2013})
\end{barticle}
\endbibitem

\bibitem[\protect\citeauthoryear{Halko et~al.}{2011}]{HMT11}
\begin{barticle}
\bauthor{\bsnm{Halko}, \binits{N.}},
\bauthor{\bsnm{Martinsson}, \binits{P.G.}},
\bauthor{\bsnm{Tropp}, \binits{J.A.}}:
\batitle{Finding structure with randomness: Probabilistic algorithms for
  constructing approximate matrix decompositions}.
\bjtitle{SIAM Review}
\bvolume{53}(\bissue{2}),
\bfpage{217}--\blpage{288}
(\byear{2011})
\end{barticle}
\endbibitem

\bibitem[\protect\citeauthoryear{Papadimitriou et~al.}{2000}]{PRT+00}
\begin{barticle}
\bauthor{\bsnm{Papadimitriou}, \binits{C.H.}},
\bauthor{\bsnm{Raghavan}, \binits{P.}},
\bauthor{\bsnm{Tamaki}, \binits{H.}},
\bauthor{\bsnm{Vempala}, \binits{S.}}:
\batitle{Latent semantic indexing: A probabilistic analysis}.
\bjtitle{Journal of Computer and Systems Sciences}
\bvolume{61}(\bissue{2}),
\bfpage{217}--\blpage{235}
(\byear{2000})
\end{barticle}
\endbibitem

\bibitem[\protect\citeauthoryear{Rokhlin et~al.}{2010}]{RST10}
\begin{barticle}
\bauthor{\bsnm{Rokhlin}, \binits{V.}},
\bauthor{\bsnm{Szlam}, \binits{A.}},
\bauthor{\bsnm{Tygert}, \binits{M.}}:
\batitle{A randomized algorithm for principal component analysis}.
\bjtitle{SIAM Journal on Matrix Analysis and Applications}
\bvolume{31}(\bissue{3}),
\bfpage{1100}--\blpage{1124}
(\byear{2010})
\end{barticle}
\endbibitem

\bibitem[\protect\citeauthoryear{Golub and Van~Loan}{1996}]{GL96}
\begin{bbook}
\bauthor{\bsnm{Golub}, \binits{G.H.}},
\bauthor{\bsnm{Van~Loan}, \binits{C.F.}}:
\bbtitle{Matrix Computations},
\bedition{3}rd edn.
\bpublisher{Johns Hopkins University Press},
\blocation{Baltimore, MD}
(\byear{1996})
\end{bbook}
\endbibitem

\bibitem[\protect\citeauthoryear{Gaa et~al.}{2025}]{GWFC25}
\begin{bchapter}
\bauthor{\bsnm{Gaa}, \binits{D.}},
\bauthor{\bsnm{Weickert}, \binits{J.}},
\bauthor{\bsnm{Farag}, \binits{I.}},
\bauthor{\bsnm{{\c{C}}i{\c{c}}ek}, \binits{{\"O}.}}:
\bctitle{Efficient representations of the diffusion echo}.
In: \beditor{\bsnm{Bubba}, \binits{T.A.}},
\beditor{\bsnm{Gaburro}, \binits{R.}},
\beditor{\bsnm{Gazzola}, \binits{S.}},
\beditor{\bsnm{Papafitsoros}, \binits{K.}},
\beditor{\bsnm{Pereyra}, \binits{M.}},
\beditor{\bsnm{Sch{\"o}nlieb}, \binits{C.-B.}} (eds.)
\bbtitle{Scale Space and Variational Methods in Computer Vision}.
\bsertitle{Lecture Notes in Computer Science},
vol. \bseriesno{15668},
pp. \bfpage{324}--\blpage{336}.
\bpublisher{Springer},
\blocation{Cham}
(\byear{2025})
\end{bchapter}
\endbibitem

\bibitem[\protect\citeauthoryear{He et~al.}{2013}]{HST12}
\begin{barticle}
\bauthor{\bsnm{He}, \binits{K.}},
\bauthor{\bsnm{Sun}, \binits{J.}},
\bauthor{\bsnm{Tang}, \binits{X.}}:
\batitle{Guided image filtering}.
\bjtitle{IEEE Transactions on Pattern Analysis and Machine Intelligence}
\bvolume{35}(\bissue{6}),
\bfpage{1397}--\blpage{1409}
(\byear{2013})
\end{barticle}
\endbibitem

\bibitem[\protect\citeauthoryear{Dam et~al.}{2003}]{DON03}
\begin{bchapter}
\bauthor{\bsnm{Dam}, \binits{E.}},
\bauthor{\bsnm{Olsen}, \binits{O.F.}},
\bauthor{\bsnm{Nielsen}, \binits{M.}}:
\bctitle{Approximating non-linear diffusion}.
In: \beditor{\bsnm{Griffin}, \binits{L.D.}},
\beditor{\bsnm{Lillholm}, \binits{M.}} (eds.)
\bbtitle{Scale Space Methods in Computer Vision}.
\bsertitle{Lecture Notes in Computer Science},
vol. \bseriesno{2695},
pp. \bfpage{117}--\blpage{131}.
\bpublisher{Springer},
\blocation{Berlin}
(\byear{2003})
\end{bchapter}
\endbibitem

\bibitem[\protect\citeauthoryear{Fischl and Schwartz}{1997}]{FS97}
\begin{barticle}
\bauthor{\bsnm{Fischl}, \binits{B.}},
\bauthor{\bsnm{Schwartz}, \binits{E.}}:
\batitle{Learning an integral equation approxiation to nonlinear anisotropic
  diffusion in image processing}.
\bjtitle{IEEE Transactions on Pattern Analysis and Machine Intelligence}
\bvolume{19}(\bissue{4}),
\bfpage{342}--\blpage{352}
(\byear{1997})
\end{barticle}
\endbibitem

\bibitem[\protect\citeauthoryear{Nitzberg and Shiota}{1992}]{NS92}
\begin{barticle}
\bauthor{\bsnm{Nitzberg}, \binits{M.}},
\bauthor{\bsnm{Shiota}, \binits{T.}}:
\batitle{Nonlinear image filtering with edge and corner enhancement}.
\bjtitle{IEEE Transactions on Pattern Analysis and Machine Intelligence}
\bvolume{14},
\bfpage{826}--\blpage{833}
(\byear{1992})
\end{barticle}
\endbibitem

\bibitem[\protect\citeauthoryear{C{\'a}rdenas et~al.}{2015}]{CWS15}
\begin{bchapter}
\bauthor{\bsnm{C{\'a}rdenas}, \binits{G.M.}},
\bauthor{\bsnm{Weickert}, \binits{J.}},
\bauthor{\bsnm{Sch{\"a}ffer}, \binits{S.}}:
\bctitle{A linear scale-space theory for continuous nonlocal evolutions}.
In: \beditor{\bsnm{Aujol}, \binits{J.-F.}},
\beditor{\bsnm{Nikolova}, \binits{M.}},
\beditor{\bsnm{Papadakis}, \binits{N.}} (eds.)
\bbtitle{Scale Space and Variational Methods in Computer Vision}.
\bsertitle{Lecture Notes in Computer Science},
vol. \bseriesno{9087},
pp. \bfpage{103}--\blpage{114}.
\bpublisher{Springer},
\blocation{Berlin}
(\byear{2015})
\end{bchapter}
\endbibitem

\bibitem[\protect\citeauthoryear{Milanfar}{2013}]{Mi13}
\begin{barticle}
\bauthor{\bsnm{Milanfar}, \binits{P.}}:
\batitle{Symmetrizing smoothing filters}.
\bjtitle{SIAM Journal on Imaging Sciences}
\bvolume{6}(\bissue{1}),
\bfpage{263}--\blpage{284}
(\byear{2013})
\end{barticle}
\endbibitem

\bibitem[\protect\citeauthoryear{Spira et~al.}{2003}]{SKS03}
\begin{bchapter}
\bauthor{\bsnm{Spira}, \binits{A.}},
\bauthor{\bsnm{Kimmel}, \binits{R.}},
\bauthor{\bsnm{Sochen}, \binits{N.}}:
\bctitle{Efficient {B}eltrami flow using a short time kernel}.
In: \beditor{\bsnm{Griffin}, \binits{L.D.}},
\beditor{\bsnm{Lillholm}, \binits{M.}} (eds.)
\bbtitle{Scale Space Methods in Computer Vision}.
\bsertitle{Lecture Notes in Computer Science},
vol. \bseriesno{2695},
pp. \bfpage{511}--\blpage{522}.
\bpublisher{Springer},
\blocation{Berlin}
(\byear{2003})
\end{bchapter}
\endbibitem

\bibitem[\protect\citeauthoryear{Pearson}{1901}]{Pe01}
\begin{barticle}
\bauthor{\bsnm{Pearson}, \binits{K.}}:
\batitle{{LIII}. {O}n lines and planes of closest fit to systems of points in
  space}.
\bjtitle{The London, Edinburgh, and Dublin Philosophical Magazine and Journal
  of Science}
\bvolume{2}(\bissue{11}),
\bfpage{559}--\blpage{572}
(\byear{1901})
\end{barticle}
\endbibitem

\bibitem[\protect\citeauthoryear{Baykova}{2016}]{Ba16}
\begin{botherref}
\oauthor{\bsnm{Baykova}, \binits{I.}}:
{PCA}-based representation of diffusion echoes.
Bachelor thesis,
Department of Computer Science, Saarland University,
Saarbr\"ucken, Germany
(2016)
\end{botherref}
\endbibitem

\bibitem[\protect\citeauthoryear{\c{C}i\c{c}ek}{2014}]{Ci14a}
\begin{botherref}
\oauthor{\bsnm{\c{C}i\c{c}ek}, \binits{{\"O}.}}:
Efficient computation and representation of the diffusion echo.
Master thesis,
Department of Computer Science, Saarland University,
Saarbr\"ucken, Germany
(2014)
\end{botherref}
\endbibitem

\bibitem[\protect\citeauthoryear{Elad}{2002}]{El02}
\begin{barticle}
\bauthor{\bsnm{Elad}, \binits{M.}}:
\batitle{On the bilateral filter and ways to improve it}.
\bjtitle{IEEE Transactions on Image Processing}
\bvolume{11}(\bissue{10}),
\bfpage{1141}--\blpage{1151}
(\byear{2002})
\end{barticle}
\endbibitem

\bibitem[\protect\citeauthoryear{{van der Vorst}}{2003}]{Vo03}
\begin{bbook}
\bauthor{\bsnm{{van der Vorst}}, \binits{H.A.}}:
\bbtitle{Iterative Krylov Methods for Large Linear Systems}.
\bpublisher{Cambridge University Press},
\blocation{Cambridge}
(\byear{2003})
\end{bbook}
\endbibitem

\bibitem[\protect\citeauthoryear{Colton}{1998}]{Co88}
\begin{bbook}
\bauthor{\bsnm{Colton}, \binits{D.}}:
\bbtitle{Partial Differential Equations}.
\bpublisher{Random House},
\blocation{New York}
(\byear{1998})
\end{bbook}
\endbibitem

\bibitem[\protect\citeauthoryear{Catt\'e et~al.}{1992}]{CLMC92}
\begin{barticle}
\bauthor{\bsnm{Catt\'e}, \binits{F.}},
\bauthor{\bsnm{Lions}, \binits{P.-L.}},
\bauthor{\bsnm{Morel}, \binits{J.-M.}},
\bauthor{\bsnm{Coll}, \binits{T.}}:
\batitle{Image selective smoothing and edge detection by nonlinear diffusion}.
\bjtitle{SIAM Journal on Numerical Analysis}
\bvolume{29}(\bissue{1}),
\bfpage{182}--\blpage{193}
(\byear{1992})
\end{barticle}
\endbibitem

\bibitem[\protect\citeauthoryear{Charbonnier et~al.}{1997}]{CBAB97}
\begin{barticle}
\bauthor{\bsnm{Charbonnier}, \binits{P.}},
\bauthor{\bsnm{Blanc--F\'eraud}, \binits{L.}},
\bauthor{\bsnm{Aubert}, \binits{G.}},
\bauthor{\bsnm{Barlaud}, \binits{M.}}:
\batitle{Deterministic edge-preserving regularization in computed imaging}.
\bjtitle{IEEE Transactions on Image Processing}
\bvolume{6}(\bissue{2}),
\bfpage{298}--\blpage{311}
(\byear{1997})
\end{barticle}
\endbibitem

\bibitem[\protect\citeauthoryear{Weickert}{1996}]{We94e}
\begin{bchapter}
\bauthor{\bsnm{Weickert}, \binits{J.}}:
\bctitle{Theoretical foundations of anisotropic diffusion in image processing}.
In: \beditor{\bsnm{Kropatsch}, \binits{W.}},
\beditor{\bsnm{Klette}, \binits{R.}},
\beditor{\bsnm{Solina}, \binits{F.}},
\beditor{\bsnm{Albrecht}, \binits{R.}} (eds.)
\bbtitle{Theoretical Foundations of Computer Vision}.
\bsertitle{Computing Supplement},
vol. \bseriesno{11},
pp. \bfpage{221}--\blpage{236}.
\bpublisher{Springer},
\blocation{Vienna, Austria}
(\byear{1996})
\end{bchapter}
\endbibitem

\bibitem[\protect\citeauthoryear{Weickert et~al.}{2013}]{WWW13}
\begin{bchapter}
\bauthor{\bsnm{Weickert}, \binits{J.}},
\bauthor{\bsnm{Welk}, \binits{M.}},
\bauthor{\bsnm{Wickert}, \binits{M.}}:
\bctitle{{L}2-stable nonstandard finite differences for anisotropic diffusion}.
In: \beditor{\bsnm{Kuijper}, \binits{A.}},
\beditor{\bsnm{Bredies}, \binits{K.}},
\beditor{\bsnm{Pock}, \binits{T.}},
\beditor{\bsnm{Bischof}, \binits{H.}} (eds.)
\bbtitle{Scale Space and Variational Methods in Computer Vision}.
\bsertitle{Lecture Notes in Computer Science},
vol. \bseriesno{7893},
pp. \bfpage{380}--\blpage{391}.
\bpublisher{Springer},
\blocation{Berlin}
(\byear{2013})
\end{bchapter}
\endbibitem

\bibitem[\protect\citeauthoryear{Schmaltz et~al.}{2014}]{SPMEWB14}
\begin{barticle}
\bauthor{\bsnm{Schmaltz}, \binits{C.}},
\bauthor{\bsnm{Peter}, \binits{P.}},
\bauthor{\bsnm{Mainberger}, \binits{M.}},
\bauthor{\bsnm{Ebel}, \binits{F.}},
\bauthor{\bsnm{Weickert}, \binits{J.}},
\bauthor{\bsnm{Bruhn}, \binits{A.}}:
\batitle{Understanding, optimising, and extending data compression with
  anisotropic diffusion}.
\bjtitle{International Journal of Computer Vision}
\bvolume{108}(\bissue{3}),
\bfpage{222}--\blpage{240}
(\byear{2014})
\end{barticle}
\endbibitem

\bibitem[\protect\citeauthoryear{Weickert and Welk}{2006}]{WW06}
\begin{bchapter}
\bauthor{\bsnm{Weickert}, \binits{J.}},
\bauthor{\bsnm{Welk}, \binits{M.}}:
\bctitle{Tensor field interpolation with {PDEs}}.
In: \beditor{\bsnm{Weickert}, \binits{J.}},
\beditor{\bsnm{Hagen}, \binits{H.}} (eds.)
\bbtitle{Visualization and Processing of Tensor Fields},
pp. \bfpage{315}--\blpage{325}.
\bpublisher{Springer},
\blocation{Berlin}
(\byear{2006})
\end{bchapter}
\endbibitem

\bibitem[\protect\citeauthoryear{Mainberger et~al.}{2011}]{MBWF11}
\begin{barticle}
\bauthor{\bsnm{Mainberger}, \binits{M.}},
\bauthor{\bsnm{Bruhn}, \binits{A.}},
\bauthor{\bsnm{Weickert}, \binits{J.}},
\bauthor{\bsnm{Forchhammer}, \binits{S.}}:
\batitle{Edge-based compression of cartoon-like images with homogeneous
  diffusion}.
\bjtitle{Pattern Recognition}
\bvolume{44}(\bissue{9}),
\bfpage{1859}--\blpage{1873}
(\byear{2011})
\end{barticle}
\endbibitem

\bibitem[\protect\citeauthoryear{Kachanov}{1959}]{Ka59}
\begin{barticle}
\bauthor{\bsnm{Kachanov}, \binits{L.M.}}:
\batitle{Variational methods of solution of plasticity problems}.
\bjtitle{Journal of Applied Mathematics and Mechanics}
\bvolume{23}(\bissue{3}),
\bfpage{880}--\blpage{883}
(\byear{1959})
\end{barticle}
\endbibitem

\bibitem[\protect\citeauthoryear{Bungert et~al.}{2023}]{BPW23}
\begin{bchapter}
\bauthor{\bsnm{Bungert}, \binits{P.}},
\bauthor{\bsnm{Peter}, \binits{P.}},
\bauthor{\bsnm{Weickert}, \binits{J.}}:
\bctitle{Image blending with osmosis}.
In: \beditor{\bsnm{Calatroni}, \binits{L.}},
\beditor{\bsnm{Donatelli}, \binits{M.}},
\beditor{\bsnm{Morigi}, \binits{S.}},
\beditor{\bsnm{Prato}, \binits{M.}},
\beditor{\bsnm{Santacesaria}, \binits{M.}} (eds.)
\bbtitle{Scale Space and Variational Methods in Computer Vision}.
\bsertitle{Lecture Notes in Computer Science},
vol. \bseriesno{14009},
pp. \bfpage{652}--\blpage{664}.
\bpublisher{Springer},
\blocation{Cham}
(\byear{2023})
\end{bchapter}
\endbibitem

\bibitem[\protect\citeauthoryear{Vogel et~al.}{2013}]{VHWS13}
\begin{bchapter}
\bauthor{\bsnm{Vogel}, \binits{O.}},
\bauthor{\bsnm{Hagenburg}, \binits{K.}},
\bauthor{\bsnm{Weickert}, \binits{J.}},
\bauthor{\bsnm{Setzer}, \binits{S.}}:
\bctitle{A fully discrete theory for linear osmosis filtering}.
In: \beditor{\bsnm{Kuijper}, \binits{A.}},
\beditor{\bsnm{Bredies}, \binits{K.}},
\beditor{\bsnm{Pock}, \binits{T.}},
\beditor{\bsnm{Bischof}, \binits{H.}} (eds.)
\bbtitle{Scale Space and Variational Methods in Computer Vision}.
\bsertitle{Lecture Notes in Computer Science},
vol. \bseriesno{7893},
pp. \bfpage{368}--\blpage{379}.
\bpublisher{Springer},
\blocation{Berlin}
(\byear{2013})
\end{bchapter}
\endbibitem

\bibitem[\protect\citeauthoryear{Horn and Johnson}{1990}]{HJ90}
\begin{bbook}
\bauthor{\bsnm{Horn}, \binits{R.A.}},
\bauthor{\bsnm{Johnson}, \binits{C.R.}}:
\bbtitle{Matrix Analysis}.
\bpublisher{Cambridge University Press},
\blocation{Cambridge, UK}
(\byear{1990})
\end{bbook}
\endbibitem

\bibitem[\protect\citeauthoryear{Bruhn et~al.}{2005}]{BWS05}
\begin{barticle}
\bauthor{\bsnm{Bruhn}, \binits{A.}},
\bauthor{\bsnm{Weickert}, \binits{J.}},
\bauthor{\bsnm{Schn\"orr}, \binits{C.}}:
\batitle{{L}ucas/{K}anade meets {H}orn/{S}chunck: Combining local and global
  optic flow methods}.
\bjtitle{International Journal of Computer Vision}
\bvolume{61}(\bissue{3}),
\bfpage{211}--\blpage{231}
(\byear{2005})
\end{barticle}
\endbibitem

\bibitem[\protect\citeauthoryear{Demetz et~al.}{2012}]{DWBZ12}
\begin{bchapter}
\bauthor{\bsnm{Demetz}, \binits{O.}},
\bauthor{\bsnm{Weickert}, \binits{J.}},
\bauthor{\bsnm{Bruhn}, \binits{A.}},
\bauthor{\bsnm{Zimmer}, \binits{H.}}:
\bctitle{Optic flow scale space}.
In: \beditor{\bsnm{Bruckstein}, \binits{A.}},
\beditor{\bsnm{Haar~Romeny}, \binits{B.}},
\beditor{\bsnm{Bronstein}, \binits{A.}},
\beditor{\bsnm{Bronstein}, \binits{M.}} (eds.)
\bbtitle{Scale Space and Variational Methods in Computer Vision}.
\bsertitle{Lecture Notes in Computer Science},
vol. \bseriesno{6667},
pp. \bfpage{713}--\blpage{724}.
\bpublisher{Springer},
\blocation{Berlin}
(\byear{2012})
\end{bchapter}
\endbibitem

\bibitem[\protect\citeauthoryear{Weickert and Schn\"orr}{2001}]{WS00b}
\begin{barticle}
\bauthor{\bsnm{Weickert}, \binits{J.}},
\bauthor{\bsnm{Schn\"orr}, \binits{C.}}:
\batitle{A theoretical framework for convex regularizers in {PDE}-based
  computation of image motion}.
\bjtitle{International Journal of Computer Vision}
\bvolume{45}(\bissue{3}),
\bfpage{245}--\blpage{264}
(\byear{2001})
\end{barticle}
\endbibitem

\bibitem[\protect\citeauthoryear{Nagel and Enkelmann}{1986}]{NE86}
\begin{barticle}
\bauthor{\bsnm{Nagel}, \binits{H.-H.}},
\bauthor{\bsnm{Enkelmann}, \binits{W.}}:
\batitle{An investigation of smoothness constraints for the estimation of
  displacement vector fields from image sequences}.
\bjtitle{IEEE Transactions on Pattern Analysis and Machine Intelligence}
\bvolume{8},
\bfpage{565}--\blpage{593}
(\byear{1986})
\end{barticle}
\endbibitem

\bibitem[\protect\citeauthoryear{Baker et~al.}{2011}]{BSLR+11}
\begin{barticle}
\bauthor{\bsnm{Baker}, \binits{S.}},
\bauthor{\bsnm{Scharstein}, \binits{D.}},
\bauthor{\bsnm{Lewis}, \binits{J.P.}},
\bauthor{\bsnm{Roth}, \binits{S.}},
\bauthor{\bsnm{Black}, \binits{M.J.}},
\bauthor{\bsnm{Szeliski}, \binits{R.}}:
\batitle{A database and evaluation methodology for optical flow}.
\bjtitle{International Journal of Computer Vision}
\bvolume{92}(\bissue{1}),
\bfpage{1}--\blpage{31}
(\byear{2011})
\end{barticle}
\endbibitem

\bibitem[\protect\citeauthoryear{Eckart and Young}{1936}]{EY36}
\begin{barticle}
\bauthor{\bsnm{Eckart}, \binits{C.}},
\bauthor{\bsnm{Young}, \binits{G.}}:
\batitle{The approximation of one matrix by another of lower rank}.
\bjtitle{Psychometrika}
\bvolume{1}(\bissue{3}),
\bfpage{211}--\blpage{218}
(\byear{1936})
\end{barticle}
\endbibitem

\bibitem[\protect\citeauthoryear{Tropp and Webber}{2023}]{TW23}
\begin{botherref}
\oauthor{\bsnm{Tropp}, \binits{J.A.}},
\oauthor{\bsnm{Webber}, \binits{R.J.}}:
Randomized algorithms for low-rank matrix approximation: Design, analysis, and
  applications.
arXiv preprint 2306.12418
(2023)
\end{botherref}
\endbibitem

\bibitem[\protect\citeauthoryear{Martinsson and Tropp}{2020}]{MT20}
\begin{barticle}
\bauthor{\bsnm{Martinsson}, \binits{P.-G.}},
\bauthor{\bsnm{Tropp}, \binits{J.A.}}:
\batitle{Randomized numerical linear algebra: Foundations and algorithms}.
\bjtitle{Acta Numerica}
\bvolume{29},
\bfpage{403}--\blpage{572}
(\byear{2020})
\end{barticle}
\endbibitem

\bibitem[\protect\citeauthoryear{Buades et~al.}{2006}]{BCM06}
\begin{barticle}
\bauthor{\bsnm{Buades}, \binits{A.}},
\bauthor{\bsnm{Coll}, \binits{B.}},
\bauthor{\bsnm{Morel}, \binits{J.-M.}}:
\batitle{Neighborhood filters and {PDE’s}}.
\bjtitle{Numerische Mathematik}
\bvolume{105}(\bissue{1}),
\bfpage{1}--\blpage{34}
(\byear{2006})
\end{barticle}
\endbibitem

\bibitem[\protect\citeauthoryear{Didas and Weickert}{2006}]{DW06}
\begin{bchapter}
\bauthor{\bsnm{Didas}, \binits{S.}},
\bauthor{\bsnm{Weickert}, \binits{J.}}:
\bctitle{From adaptive averaging to accelerated nonlinear diffusion filtering}.
In: \beditor{\bsnm{Franke}, \binits{K.}},
\beditor{\bsnm{M\"uller}, \binits{K.-R.}},
\beditor{\bsnm{Nickolay}, \binits{B.}},
\beditor{\bsnm{Sch\"afer}, \binits{R.}} (eds.)
\bbtitle{Pattern Recognition}.
\bsertitle{Lecture Notes in Computer Science},
vol. \bseriesno{4174},
pp. \bfpage{101}--\blpage{110}.
\bpublisher{Springer},
\blocation{Berlin}
(\byear{2006})
\end{bchapter}
\endbibitem

\bibitem[\protect\citeauthoryear{Sochen et~al.}{2001}]{SKB01}
\begin{barticle}
\bauthor{\bsnm{Sochen}, \binits{N.}},
\bauthor{\bsnm{Kimmel}, \binits{R.}},
\bauthor{\bsnm{Bruckstein}, \binits{F.}}:
\batitle{Diffusions and confusions in signal and image processing}.
\bjtitle{Journal of Mathematical Imaging and Vision}
\bvolume{14}(\bissue{3}),
\bfpage{195}--\blpage{210}
(\byear{2001})
\end{barticle}
\endbibitem

\bibitem[\protect\citeauthoryear{Paris et~al.}{2009}]{PKTD09}
\begin{barticle}
\bauthor{\bsnm{Paris}, \binits{S.}},
\bauthor{\bsnm{Kornprobst}, \binits{P.}},
\bauthor{\bsnm{Tumblin}, \binits{J.}},
\bauthor{\bsnm{Durand}, \binits{F.}}:
\batitle{Bilateral filtering: Theory and applications}.
\bjtitle{Foundations and Trends in Computer Graphics and Vision}
\bvolume{4}(\bissue{1}),
\bfpage{1}--\blpage{73}
(\byear{2009})
\end{barticle}
\endbibitem

\bibitem[\protect\citeauthoryear{Jennewein}{2014}]{Je14}
\begin{botherref}
\oauthor{\bsnm{Jennewein}, \binits{S.}}:
Interpretation nichtlinearer bildverarbeitungsmethoden anhand ihres
  filterechos.
High school teacher thesis,
Department of Computer Science, Saarland University,
Saarbr\"ucken, Germany
(2014).
In German
\end{botherref}
\endbibitem

\bibitem[\protect\citeauthoryear{Belhachmi et~al.}{2009}]{BBBW09}
\begin{barticle}
\bauthor{\bsnm{Belhachmi}, \binits{Z.}},
\bauthor{\bsnm{Bucur}, \binits{D.}},
\bauthor{\bsnm{Burgeth}, \binits{B.}},
\bauthor{\bsnm{Weickert}, \binits{J.}}:
\batitle{How to choose interpolation data in images}.
\bjtitle{SIAM Journal on Applied Mathematics}
\bvolume{70}(\bissue{1}),
\bfpage{333}--\blpage{352}
(\byear{2009})
\end{barticle}
\endbibitem

\bibitem[\protect\citeauthoryear{Anderson et~al.}{1999}]{ABB+99}
\begin{bbook}
\bauthor{\bsnm{Anderson}, \binits{E.}},
\bauthor{\bsnm{Bai}, \binits{Z.}},
\bauthor{\bsnm{Bischof}, \binits{C.}},
\bauthor{\bsnm{Blackford}, \binits{S.}},
\bauthor{\bsnm{Demmel}, \binits{J.}},
\bauthor{\bsnm{Dongarra}, \binits{J.}},
\bauthor{\bsnm{Du~Croz}, \binits{J.}},
\bauthor{\bsnm{Greenbaum}, \binits{A.}},
\bauthor{\bsnm{Hammarling}, \binits{S.}},
\bauthor{\bsnm{McKenney}, \binits{A.}},
\bauthor{\bsnm{Sorensen}, \binits{D.}}:
\bbtitle{{LAPACK} Users' Guide},
\bedition{3}rd edn.
\bpublisher{Society for Industrial and Applied Mathematics},
\blocation{Philadelphia, PA}
(\byear{1999})
\end{bbook}
\endbibitem

\bibitem[\protect\citeauthoryear{Hutchinson}{1990}]{Hu90}
\begin{barticle}
\bauthor{\bsnm{Hutchinson}, \binits{M.F.}}:
\batitle{A stochastic estimator of the trace of the influence matrix for
  {L}aplacian smoothing splines}.
\bjtitle{Communications in Statistics - Simulation and Computation}
\bvolume{19}(\bissue{2}),
\bfpage{433}--\blpage{450}
(\byear{1990})
\end{barticle}
\endbibitem

\bibitem[\protect\citeauthoryear{Silver and R\"oder}{1997}]{SR97}
\begin{barticle}
\bauthor{\bsnm{Silver}, \binits{R.N.}},
\bauthor{\bsnm{R\"oder}, \binits{H.}}:
\batitle{Calculation of densities of states and spectral functions by chebyshev
  recursion and maximum entropy}.
\bjtitle{Physical Review E}
\bvolume{56}(\bissue{4}),
\bfpage{4822}--\blpage{4829}
(\byear{1997})
\end{barticle}
\endbibitem

\bibitem[\protect\citeauthoryear{Avron and Toledo}{2011}]{AT11}
\begin{barticle}
\bauthor{\bsnm{Avron}, \binits{H.}},
\bauthor{\bsnm{Toledo}, \binits{S.}}:
\batitle{Randomized algorithms for estimating the trace of an implicit
  symmetric positive semi-definite matrix}.
\bjtitle{Journal of the ACM}
\bvolume{58}(\bissue{2}),
\bfpage{1}--\blpage{34}
(\byear{2011})
\end{barticle}
\endbibitem

\end{thebibliography}

\end{document}